\documentclass[nofootinbib, aps]{revtex4}
\usepackage{amstext,amsmath,amssymb}
\usepackage[dvips]{graphicx}
\usepackage{latexsym}
\newcommand{\N}{\mathbb{N}}

\newcommand{\C}{\mathbb{C}}
\newcommand{\SU}{\mathrm{SU}}
\newcommand{\U}{\mathrm{U}}
\renewcommand{\a}{\alpha}
\newcommand{\ba}{\bar{\a}}

\renewcommand{\d}{\delta}
\newcommand{\g}{\gamma}
\newcommand{\bg}{\bar{\g}}
\newcommand{\tg}{\tilde{\g}}
\newcommand{\G}{\Gamma}
\newcommand{\D}{\Delta}
\renewcommand{\l}{\lambda}
\renewcommand{\L}{\Lambda}
\newcommand{\m}{\mu}
\renewcommand{\o}{\theta}
\newcommand{\p}{\phi}
\renewcommand{\P}{\Phi}
\newcommand{\s}{\sigma}
\def\rar{\rightarrow}
\newcommand{\rep}[1]{D^{j_{#1}}_{m_{#1}n_{#1}}(g_{#1})}
\newcommand{\dk}[1]{\d(\bar{g}_{#1}\bar{\a}^{-1}\a g_{#1}^{-1})}
\newcommand{\dvp}[3]{\d(\bar{g}_{#1}\a^{-1}_{#2}\a_{#3}g_{#1}^{-1})}
\newcommand{\dvc}[4]{\d(\bar{g}_{#1}\a^{-1}_{#2}\a_{#3}u_{#4}^{-1}hu_{#4}g_{#1}^{-1})}

\newcommand{\dvg}[3]{\d(\bar{g}_{#1}\a^{-1}_{#2}\g_{#3#2}\a_{#3}g_{#1}^{-1})}
\newcommand{\cK}{\mathcal{K}}
\newcommand{\cP}{\mathcal{P}}
\newcommand{\cV}{\mathcal{V}}
\newcommand{\cD}{\mathcal{D}}
\newcommand{\be}{\begin{equation}}
\newcommand{\ee}{\end{equation}}
\def\beq{\begin{eqnarray}}
\def\eeq{\end{eqnarray}}
\begin{document}
%
%
%
%
%
%
%
%
%
%
\title{Group field theory formulation of 3d quantum gravity coupled to matter fields}
\author{{\bf Daniele Oriti\footnote{d.oriti@damtp.cam.ac.uk} and James Ryan\footnote{j.p.ryan@damtp.cam.ac.uk}}}
\vspace{0.5cm} \affiliation{Department of Applied Mathematics and
Theoretical Physics, \\ Centre for Mathematical Sciences, \\
University of Cambridge, Cambridge CB3 0WA, England, EU}
\begin{abstract}
\begin{center}
{\small ABSTRACT}
\end{center}
We present a new group field theory describing 3d Riemannian quantum gravity
coupled to matter fields for any choice of spin and mass. The
perturbative expansion of the partition function produces fat graphs
colored with $SU(2)$ algebraic data,
from which one can reconstruct at once a 3-dimensional simplicial
complex representing spacetime and its geometry, like in the
Ponzano-Regge formulation of pure 3d quantum gravity, and the Feynman
graphs for the matter fields. The model then assigns quantum
amplitudes to these fat graphs given by spin foam models for gravity
coupled to interacting massive spinning point particles, whose
properties we discuss.
\end{abstract}

\maketitle
%
%
%
%
%
%
%
%
%
%
\section{Introduction}

\subsection{Background}
Spin foam models \cite{daniele, alex} emerged recently as a
general formalism for quantum gravity, and a point of convergence
of different approaches, including loop quantum gravity (of which
they may be thought of as a path integral formulation),
topological field theories using ideas from category theory, and
simplicial gravity. They assign geometric data to the simplicial
spacetime in an algebraic form. In turn, spin foam models have
been shown to be obtainable from so-called group field theories
\cite{iogft,laurentgft}, i.e. field theories over group manifolds
that can be seen as a generalisation of matrix models in that they
produce, in their perturbative expansion in Feynman graphs, a sum
over simplicial complexes of dimension higher than 2, with the
configuration/momentum variables of the field being interpreted as
geometric data for these complexes. This sum over simplicial
complexes constructed as Feynman diagrams of the group field
theory, implies also a sum over topologies; therefore the group
field theory formalism can be interpreted \cite{laurentgft,iogft}
as a realization of a third quantization of gravity at the
simplicial level. There exist promising and currently much studied
spin foam models, and group field theories, in 4-dimensions
\cite{daniele, alex}, whose validity is, however, still under
investigation. In the simpler case of 3-dimensional gravity (both
Riemannian and Lorentzian, with and without cosmological constant)
it is now established that spin foam models provide a consistent
quantization, equivalent, but also presenting distinctive
advantages, to those obtained from other approaches. The relevant
model for 3d gravity without cosmological constant is the
Ponzano-Regge model \cite{laurentPRI}, whose group field theory
derivation was given by Boulatov in \cite{boulatov}.

\medskip

The coupling of matter fields to quantum gravity in the spin foam
framework is of paramount importance for various reasons, apart
from the obvious one that for a consistent theory of quantum
gravity to be correct, it should be able to describe in full the
interactions between gravity and matter fields. First of all,
matter coupling may provide the best, if not the only, way to
define quantum observables for the theory that have a clear
physical meaning, given that such observables are very difficult
to define in a pure gravity theory \cite{carloObs, carlopartial,
bianca}. In particular, the inclusion of matter fields may prove
to be the main avenue towards the construction of a quantum
gravity phenomenology that could be put to test in future
experiments \cite{amelino}, the idea being that quantum gravity
will modify the usual dynamics of matter fields (e.g. dispersion
relations, scattering amplitudes, etc.) even in an approximately
flat background, leading to potentially testable effects. Also, it
is hoped that quantum gravity will not only modify the usual
predictions of quantum field theory, but also solve various
problems of the same, including that of ultraviolet divergences,
providing a kind of built-in covariant cut-off at the Planck
scale. Whether any of these hopes are actually fulfilled can be
shown only by explicit work on matter couplings in quantum gravity
models, including spin foams. Recently much research has been
devoted to this issue. After some more speculative proposals
\cite{crane, lee, mikovic}, a spin foam model for gravity coupled
to gauge fields in 4d has been constructed in \cite{danhend}, and
later re-derived using different methods in \cite{mikovicgauge},
but most of the work on matter fields in spin foam models for
quantum gravity has been done in the past two years
\cite{KarimAlex,barrettfeynman,laurentPRI,laurentPRII,laurentPRIII}
and focused on the 3d case. The work of \cite{KarimAlex}, starts
off from a canonical perspective and build on results that have
been obtained in the context of loop quantum gravity
\cite{carlo-hugo,kirill,kirill-john,thomas}, and obtains a spin
foam description of the dynamics of matter and quantum gravity by
an explicit construction of the projector onto physical quantum
states of the coupled system. On the other hand the construction
of \cite{laurentPRI,laurentPRII,laurentPRIII} uses a
covariant/path integral picture from the start and is phrased in a
discrete (simplicial) context. It is then more directly linked
both in terms of language used and results obtained to the group
field theory context. In fact, in this paper we are going to
construct a group field theory whose corresponding Feynman
amplitudes are exactly the spin foam amplitudes for gravity
coupled to particles of any spin derived in \cite{laurentPRI} and
further analysed in \cite{laurentPRII,
  laurentPRIII}.
The basic ideas of the construction in \cite{laurentPRI} are the
following: 1) Feynman graphs for particle interactions (like those
coming from usual quantum field theories), including their coupling with gravity, can be considered as
(non-local) observables for quantum gravity, and therefore treated
as such, i.e. inserted as appropriate operators in the pure
gravity partition function, so that the theory would allow the
computation of their expectation value; the formal expression of
this expectation value is that of a modified spin foam model where
the modified amplitudes encode both the geometric and particle
degrees of freedom; 2) matter arises as a kind of
symmetry-breaking singular configuration of the gravitational
field, in the sense that pure gauge degrees of freedom are turned
into physical degrees of freedom (characterizing matter) at the
location of the particles. We will see in the following how these
ideas are realised also in our model at the group field theory
level. In particular, the first idea has a very natural
implementation in the group field theory formalism (in a sense, it
suggests such a formalism) and it is rather new and likely to be
a key for future developments towards quantum gravity
phenomenology, thanks to the possibility of extracting an
effective non-commutative field theory for matter fields encoding
the quantum gravity effects \cite{laurentPRIII}. On the other
hand, the idea of matter as a topological defect of gravity in 3d
is well-established since the work of S. Deser, R. Jackiw, and G.
't Hooft \cite{DJT,matschullwelling} both at the classical and
quantum continuum level, but finds a beautiful purely algebraic and
combinatorial realisation in the spin foam construction of
\cite{laurentPRI}, and therefore in the present work. A similar
use of Feynman graphs as observables for quantum gravity, as a
basis for studying the coupling of matter fields to it, was done
in \cite{barrettfeynman} in the context of the Turaev-Viro spin
foam model for 3d gravity with cosmological constant, where the
diagrams considered included colored knots. It is of great
importance for the group field theory programme
\cite{laurentgft,iogft} to be able to include matter couplings in
it, and reproduce the known coupled spin foam models, in
order to be entitled to consider it a fundamental definition of a
theory of quantum gravity, and in particular as the truly fundamental definition of spin foam models themselves, rather than just an auxiliary formalism.
Work on a group field theoretic description of quantum gravity
coupled to matter fields has started only very recently, the first
constructions having been presented in \cite{kirillgft} and
\cite{us}. In \cite{kirillgft} a very elegant extension of the
Boulatov group field theory model for 3d quantum gravity to the
$DSU(2)$ quantum group is performed, motivated by the fact that
the coupled spin foam amplitudes of \cite{laurentPRI} present a
symmetry
  under this particular deformation of the Poincare' group
  \cite{laurentPRII} and that particle states result in being labeled
  by $DSU(2)$ representations. However, the resulting model does not
  have a clear interpretation in terms of particle
  configurations. Still in \cite{kirillgft} a new class of spin foam
  models admitting such interpretation is constructed using chain mail
  techniques (see also \cite{laurentPRII}), but with no derivation of
  the same from group field theories. In \cite{us} a group field
  theory that produces spin foam amplitudes for matter coupled to 3d
  gravity was proposed, and shown to reproduce the amplitudes of
  \cite{laurentPRI} in the special case of massive scalar
  particles/fields (more precisely, for particles with no spin nor
  angular momentum). The model we present in this paper can be seen as
  a generalisation and an appropriate modification of that proposed in
  \cite{us}; it does not only generalise it to the case of spinning
  matter fields and generic interaction, but it also uses a simpler
  formalism with a clearer physical interpretation, as we will show in
  the following. This is not a direct generalisation, however, in the sense that even in the particular case of spinless particles it provides an {\it alternative} way to couple matter to gravity in group field theory, as we will see in the following. We believe that the model in
\cite{us} has a very interesting structure and that its peculiar features deserve further investigation.

\subsection{The new GFT model: general ideas}
Let us summarise and introduce the main ideas behind our model. Of
course, all of the following will be made precise and explicit
later in this work. The spin foam model of \cite{laurentPRI}, as
we mentioned above, is based (as is the work of
\cite{barrettfeynman}) on the idea that one should couple a full
history of particle interactions and evolution, represented by a
Feynman diagram of the type produced by matter field theories, to
a history of the gravitational field represented by a spin foam,
and define an appropriate quantum amplitude for the composite
history. Now, the spin foam itself is just a specific Feynman
diagram for the group field theory so that the coupled spin foam
is best interpreted as a Feynman diagram for both gravity and
matter fields and this leads naturally to the search for the
modified group field theory that generates it in its perturbative
expansion. This means that we want to realise the third
quantization of gravity in a simplicial setting and the second
quantization of matter living on the same simplicial structures in
one stroke, and define a field theory on a group manifold that
produces at once, in perturbative expansion, a sum over spacetimes
and geometries and a sum over Feynman diagrams for matter
interactions, understood as taking place in such spacetimes. As
for the representation of
  matter degrees of freedom, we base ourselves on the extensive
  knowledge of matter in 3d gravity coming from both continuum
  classical analyses \cite{DJT}, loop quantum gravity \cite{KarimAlex}
  and spin foams \cite{laurentPRI}. By virtue of its spin (and angular
  momentum), a particle
  breaks locally the gauge invariance of pure gravity, so that the
  $SU(2)$ gauge degrees of freedom are turned into
  physical degrees of freedom of the particle at its location. This is a kinematical feature that is evident before any
  dynamics are imposed. In the framework of loop quantum gravity, and
  thus for what concerns the boundary states of spin foam models, this
  implies that the relevant combinatorial/algebraic structures
  describing states of the gravity+matter system are {\it open spin
    networks} with one loose end labelled by the spin of the particle;
  one can thus encode the presence of a particle in the spin network
  describing a boundary state of a spin foam model by replacing one of
  its 3-valent vertices (in 3d) with a 4-valent vertex having an extra
  loose link labelled with the particle data. In the dual simplicial
  geometric interpretation of spin networks in a spin foam context,
  this means that one is replacing a triangle on the boundary of the
  simplicial 3-manifold dual to the spin foam, with a modified
  triangle carrying a particle and thus possessing a modified geometry. In the simplicial third quantization provided
  by the group field theory the object that corresponds to a geometric
  triangle is the field itself, so the starting idea for
  extending the setting to include spinning particles is to allow for
  a different type of field with a modified combinatorial structure
  and a modified gauge symmetry, as we will see. The mass of the
  particles can be dealt with at the dynamical level, since it does
  not influence the kinematics of the fields, and should manifest
  itself by producing locally a deviation from the pure gravity flatness
  constraint at the location of the particle, i.e. as the presence of
  distributional curvature or a conical singularity along the particle trajectories. In other words, we are going to encode the presence
  of a massive particle in its {\it interaction} with the
  gravitational field.  So we modify the interaction term of the
  pure gravity group field theory to include extra terms describing
  both the interaction of matter fields among themselves (modified by the gravity degrees of freedom), {\it and} the
  propagation of these matter fields in the quantum gravity
  background, effected by the
  generation of curvature along the particle trajectory, again
  following \cite{laurentPRI}.

The end result is a field theory whose perturbative expansion
gives fat graphs, as in the pure gravity case, but  with extra
combinatorial structures and algebraic data, so that one can
identify from them both a labelled 2-complex that allows to
reconstruct a 3d triangulation, its geometry {\it and} a Feynman
graph for the interacting matter fields (of any spin and mass)
alongside it. The quantum amplitudes for these fat graphs then
describe how the matter field Feynman graphs are embedded in the
simplicial complex and assign a total probability amplitude to the
gravity plus matter configuration, that coincide with the
amplitudes constructed in \cite{laurentPRI}.
%
%
%
%
%
%
%
%
%
%
\section{The model}
\label{model} To recapitulate, the main aim for this section is to
construct a group field theory with the following hallmark: when
we expand its partition function, the sum over Feynman diagrams
contains a sum over spin foams for matter coupled to gravity.
These spin foam amplitudes were obtained recently from a path
integral quantisation of first order gravity coupled to
relativistic point particles (with arbitrary spin)
\cite{laurentPRI}.  Results and properties of this model are
outlined in Appendix \ref{cpr},  and we will refer to these spin
foam amplitudes as Coupled Ponzano-Regge (CPR) amplitudes.

In conventional field theory, we usually distinguish between kinematic (field and symmetries)
and dynamic structures (action, partition function, etc). We follow a similar route here.

The outline of this section is as follows: we begin by defining
the fields and their symmetries and show how they incorporate the
kinematic scene;  later, we define an action and examine its
dynamic input; finally, we provide a more in-depth discussion of
generic Feynman amplitudes to link back to the spin foam stage.
\subsection{Classical and quantum kinematics}
\label{ckin} We divide our exploration of the kinematic sector
into two parts: pure gravity and matter coupled to gravity, dealing with both in a similar way.

We describe the pure gravity sector of our theory using the field
arising in Boulatov's field theory \cite{boulatov}. This theory is
a GFT for 3d Riemannian quantum gravity: its Feynman diagrams are
Ponzano-Regge (PR) spin foams associated to 3-(pseudo)manifolds.
The sum in the expansion of the partition function is over all
geometries for given topology and chosen triangulation (that is,
over all triangulations for given topology), as well as over all
topologies. Since the PR amplitude is a topological invariant it
does not depend on the geometry. First of all, the field is
defined as a map from the Cartesian product of three copies of
$\SU(2)$ to the complex numbers:
\begin{equation}\p:\SU(2)\times\SU(2)\times\SU(2)\rar\C;\quad\p(g_1,g_2,g_3). \label{pdef}\end{equation}
We refer to this field as the Boulatov field. It has two
symmetries:
\begin{itemize} \item we require $\p$ to be invariant
under (even) elements $\s$ of the permutation group of three
elements $S_3$, acting on the field variables,
\begin{equation}\p(g_1,g_2,g_3)=\p(g_{\s(1)},g_{\s(2)},g_{\s(3)}),\label{pperm}\end{equation}
the Feynman amplitudes produced by the corresponding group field
theory are in one-to-one correspondence with {\it orientable}
2-complexes, as explained in \cite{dfkr, dep}; odd permutations
map this field to its complex conjugate;
\item we require the field to be Lorentz invariant; we ensure this by projecting the field onto its $\SU(2)$ invariant part:
\begin{equation}P_{\a}\p(g_1,g_2,g_3)\equiv\int_{\SU(2)}d\a\;\p(g_1\a,g_2\a,g_3\a).\label{plor}\end{equation}
\end{itemize}
Now that we have defined the field and its symmetries we want to
recover a full understanding of how these structures relate to the
kinematic regime. We then expand this field into its Fourier modes
by means of a Peter-Weyl decomposition. This gives a field
dependence on three irreducible representations, one corresponding
to each copy of $\SU(2)$.  We perform the expansion on the
projected field in Appendix \ref{exp} explicitly and the result
is: \begin{equation}
P_{\a}\p(g_1,g_2,g_3)=\sum_{\substack{j_i,m_i,n_i \\ 1\leq i\leq
3}}\sqrt{d_{j_1}d_{j_2}d_{j_3}}\P^{j_1\;\;j_2\;\;j_3}_{m_1m_2m_3}\rep{1}\rep{2}\rep{3}C^{j_1\;j_2\,j_3}_{n_1n_2n_3},\label{pexp}\end{equation}
where $D$ is a representation matrix of $\SU(2)$, $C$ is an
$\SU(2)$ $3j$-symbol and $d_j$ is the dimension of the $j$
representation. Furthermore, the field $\P$ has the symmetries of
an $\SU(2)$ $3j$-symbol\footnote{The $3j$-symbol is the unique
invariant intertwiner of three representations of $\SU(2)$.}.

The group field theory formalism is a path integral formulation
and transition amplitudes arise as \lq two-point' functions where
the two \lq points' are possible boundary data of the theory. The
states have the same structure as spin network functionals. These
are \lq coloured' closed trivalent graphs, where the edges are
labelled by matrix elements $D^{j}_{mn}(g)$ of the holonomy $g$
along the edge in a fixed representation $j$ and the vertices are
labelled by invariant intertwiners $C^{j_1\;j_2\,j_3}_{n_1n_2n_3}$
contracting the matrices. Thus a single vertex and its three
incident edges are given by
\begin{equation}\rep{1}\rep{2}\rep{3}C^{j_1\;j_2\,j_3}_{n_1n_2n_3},\label{pvert}\end{equation}
which we recognise from (\ref{pexp}) as constituting much of the
Boulatov field.  Indeed, we seek a way to translate the spin
network functional into the language of group field theory. We
accomplish this naturally by considering the same spin network
graphs but with the edges now labelled by the fixed representation
and the vertices by $\P^{j_1\;\;j_2\;\;j_3}_{m_1m_2m_3}$ which is
indeed an invariant intertwiner.  Thus a boundary state in the
Boulatov model is a product of $\P$ fields \cite{reisenberger},
contracted with respect to the \lq magnetic' indices $m$, and
diagrammatically represented by a spin network. This state charts
the gravitational information of a 2d hypersurface.
\begin{figure}[h]
\begin{center}
\includegraphics{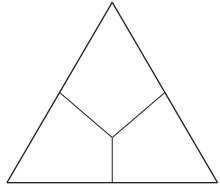}
\end{center}
\caption{\label{pfield} The pure gravity field.}
\end{figure}

As for the geometric interpretation, in
the kinematic arena, a spin network vertex is dual topologically
to a triangle, the representations are related to its edge lengths and the intertwiner ensures that the triangle
inequalities are satisfied; therefore, we think of the Boulatov
field as representing a triangle in the triangulation of a
boundary 2-manifold kinematically, and later of a 3-manifold
dynamically.

\medskip
There is a new field to represent the sector of our theory where
matter couples to gravity. This field
provides us with explicit information about the momentum and spin
of the particle.  These are the two quantities needed to
thoroughly account for a point particle coupled to gravity.  We
define the field as a map from the Cartesian product of four
copies of $\SU(2)$ to the complex numbers:
\begin{equation}\psi_s(g_1,g_2,g_3;u):\SU(2)\times\SU(2)\times\SU(2)\times\SU(2)\rar\C. \end{equation}
We refer to $\psi_s$ as the coupled field. It has four arguments,
the first three still encapsulate the gravitational degrees of
freedom while the new fourth argument is related to the momentum
of the particle.
We perform a partial decomposition to clarify the
definition of the field.
\begin{equation}
\qquad\psi_s(g_1,g_2,g_3;u)=\sum_{I,n}\psi^{I}_{sn}(g_1,g_2,g_3)D^{I}_{sn}(u).\label{cdef}\end{equation}
The $s$ index is fixed and refers to the spin of the particle,
the representation $I$ denotes its total angular momentum.

Also, the new field $\psi_s$ can be understood as the result of
projecting a generic four-argument field $\psi$ to a specific spin
component $s$ by means of a projector operator $P_s$ acting, on its
fourth argument, as:

\begin{equation}
\psi_s(g_1,g_2,g_3;u) = (P_s \psi) (g_1,g_2,g_3;u)=\sum_J \int_{SU(2)}
dg d_J D^J_{ss}(ug^{-1} \psi (g_1,g_2,g_3;g) .
\end{equation}

We impose one symmetry on the field.  This is Lorentz symmetry that is once
again ensured by projection onto the $\SU(2)$ invariant part of
the field, by simultaneous right action on all four arguments:
\begin{equation}P_{\a}\psi_s(g_1,g_2,g_3;u)\equiv\int_{\SU(2)}d\a\;\psi_s(g_1\a,g_2\a,g_3\a;u\a).\label{clor}\end{equation}
We will discuss later why we do not impose permutation symmetry
and what consequences this has on the resulting Feynman
amplitudes.

Once again we expand this field into its Fourier modes:
\begin{equation}
P_{\a}\psi_s(g_1,g_2,g_3;u)=\sum_{\substack{I,n,j_i,m_i,n_i\\
1\leq i\leq
3}}\sum_{\L}\sqrt{d_{j_1}d_{j_2}d_{j_3}d_I}\Psi^{j_1\;\;j_2\;\;j_3\;\,I\;\L}_{m_1m_2m_3s}
\rep{1}\rep{2}\rep{3}D^{I}_{sn}(u)\tilde{C}^{j_1\;j_2\,j_3\,I\,\L}_{n_1n_2n_3n},\label{cexp}\end{equation}
where $\tilde{C}$ is an invariant intertwiner of four
representations, $\L$ labels a basis in the space of 4-valent
intertwiners\footnote{The vector space of 4-valent intertwiners is
not 1-dimensional, unlike the trivalent case.} and $\Psi$ has the
symmetries of a 4-valent $SU(2)$ intertwiner.

We use our knowledge of the boundary states of the CPR spin foam model to
explain the kinematic information contained in the coupled field.  The
boundary spin foam states for matter coupled to gravity are open
coloured trivalent spin network functionals.  That is, some edges
do not join trivalent vertices but instead join a trivalent vertex
to an endpoint.  On these edges and endpoints reside the kinematic
data of the particles.  The edges are labelled with the matrix
elements of the holonomy along that edge and at the endpoints are
projected down to the spin-$s$ component.  The matrix elements
naturally furnish the particle arena with a Poincar\'e
representation labelled by its spin and mass $(s,m)$:
\begin{equation} V_{s,m}=\bigoplus_I\left\{D^{I}_{sn}(u):I-s\in\N\;;\;|n|\leq
I\right\}.\label{vect}\end{equation} In the Poincar\'e
representation of a particle, $I$ labels the particle's total
angular momentum, $s$ its spin, and $u$ its momentum.  Therefore,
we label a trivalent vertex with an open edge and two normal edges
with
\begin{equation}\rep{1}\rep{2}D^{I}_{sn}(u)C^{j_1\;j_2\,I}_{n_1n_2n}.\label{cvert}\end{equation}

\begin{figure}[h]
\begin{center}
\includegraphics{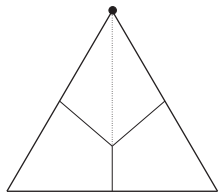}
\end{center}
\caption{\label{cfield} The coupled field.}
\end{figure}

To translate this state into the GFT language we want to consider
the same open trivalent spin network and label the edges with
representations and the trivalent vertices with fields, as we did
in the pure gravity case.  For those vertices with three normal
edges, we can label them as before with the Boulatov field.  For
those special vertices with an open edge, we wish to label them
with the coupled field
$\Psi^{j_1\;\;j_2\;\;j_3\;I\,\L}_{m_1m_2m_3s}$; this naturally
labels a 4-valent vertex, which in turn can be decomposed into two
trivalent vertices joined by a new intermediate edge that is
labelled by the $\SU(2)$ representation $\L$.  Thus we use the
coupled field to label two trivalent vertices: one normal and one
special. In the end, a boundary state is given by a product of
$\P$ and $\Psi$ fields, contracted according to the combinatorics
of the graph. We view the $\psi_s$ field as a triangle with extra
particle degrees of freedom on one of its vertices, as seen in
FIG.\ref{cfield}.

\subsection{Classical dynamics and kinetic and vertex terms}
\label{cdyn} The action defines the classical dynamics of the
fields.  From this, one could calculate the classical equations of
motion of the group field theory (i.e. Euler-Lagrange equations).
The classical equations of motion of the Boulatov GFT describe the
{\it local} evolution of pure quantum gravity in a simplicial
setting, and a recent proposal by Freidel \cite{laurentgft},
suggests that the classical structure of the GFT is what only
matters for the definition of the inner product between canonical
states in a GFT formulation of pure quantum gravity.  Certainly, a
similar interpretation is possible for our coupled GFT, and a
parallel analysis should be carried out for the GFT model we
propose here, so to unveil all the information about local
evolution and canonical inner product for gravity coupled to
matter, but we leave this for future work. We state our action as
follows:
\begin{equation}\begin{split} S[\p,\psi_s]= &\; S_{gr} + S_{gr+mat} = S_{gr} + S_{gr+mat}^{kin} + S_{gr+mat}^{int-2}+ S_{gr+mat}^{int-3} +
S_{gr+mat}^{int-4}=\\ &\;=\frac{1}{2}\int
\prod^{3}_{i=1}dg_i\;P_{\a}\p(g_1,g_2,g_3)P_{\bar{\a}}\p(g_1,g_2,g_3)\\
&+\frac{\l}{4!}\int\prod^{6}_{i=1}dg_i\;P_{\a_1}\p(g_1,g_2,g_3)P_{\a_2}\p(g_3,g_5,g_4)P_{\a_3}\p(g_4,g_2,g_6)P_{\a_4}\p(g_6,g_5,g_1)\\
&+\frac{1}{2}\int\prod^{3}_{i=1} dg_i\, du\;
P_{\a}\psi_{s}(g_1,g_2,g_3;u)P_{\bar{\a}}\psi_{s}(g_1,g_2,g_3;u)\\
&+\mu_2\int \prod^{6}_{i=1}dg_i\, du_a\;
P_{\a_1}\psi_{s}(g_1,g_2,g_3u_a^{-1}hu_a;u_a)P_{\a_2}\psi_{s}(g_4u_a^{-1}h^{-1}u_a,g_3,g_5;u_a)P_{\a_3}\p(g_4,g_2,g_6)P_{\a_4}\p(g_6,g_5,g_1)\\
&+\mu_3\int \prod^{6}_{i=1}dg_i\, du_a\, du_b\, du_c\;
P_{\a_1}\psi_{s}(g_1,g_2,g_3u_a^{-1}hu_a;u_a)P_{\a_2}\psi_{s}(g_4u_b^{-1}h^{-1}u_b,g_3,g_5;u_b)P_{\a_3}\psi_{s}(g_6,g_4,g_2u_c^{-1}hu_c;u_c)\\
&\phantom{xxxxxxxxxx}\times
P_{\a_4}\p(g_6,g_5,g_1)\d(u_a^{-1}hu_au_b^{-1}h^{-1}u_bu_c^{-1}hu_c)\sum_{\substack{I_a,I_b,I_c\\
n_a,n_b,n_c}}
D^{I_a}_{sn_a}(u_a)D^{I_b}_{sn_b}(u_b)D^{I_c}_{sn_c}(u_c)C^{I_a\,I_b\,I_c}_{n_an_bn_c}\\
&+\mu_4\int\prod^{6}_{i=1}dg_i\,  du_a\, du_b\, du_c\, du_d\;
P_{\a_1}\psi_{s}(g_1,g_2,g_3u_a^{-1}hu_a;u_a)P_{\a_2}\psi_{s}(g_4u_b^{-1}h^{-1}u_b,g_3,g_5;u_b)\\
&\phantom{xxxxxxxxxx}\times
P_{\a_3}\psi_{s}(g_2u_d^{-1}hu_d,g_6,g_4;u_c)P_{\a_4}\psi_{s}(g_6,g_4,g_2;u_d)\d(u_a^{-1}hu_au_b^{-1}h^{-1}u_bu_ch^{-1}u_cu_d^{-1}hu_d)\\
&\phantom{xxxxxxxxxx}\times\sum_{\substack{I_a,I_b,I_c,I_d,\L\\
n_a,n_b,n_c,n_d}}
D^{I_a}_{sn_a}(u_a)D^{I_b}_{sn_b}(u_b)D^{I_c}_{sn_c}(u_c)D^{I_d}_{sn_d}(u_d)\tilde{C}^{I_a\,I_b\,I_c\,I_d\,\L}_{n_an_bn_cn_d},
\label{action}\end{split}\end{equation} where $h\in U(1)$ encodes
the mass of the particle\footnote{$h=e^{mJ_0}$ and $J_0$ is an
su(2) generator of the U(1) subgroup.}. Furthermore, the first
line of the equation gives a schematic form of the later terms.
$S_{gr}$ symbolises the first two terms (pure gravity), and
$S_{gr+mat}$ binds up the later four terms (matter coupled to
gravity). We give each of the matter terms its own separate name,
for example $S_{gr+mat}^{int-2}$ denotes a vertex term with a
bivalent particle interaction.  We explain this in more detail
later.

We deal in Section \ref{qdyn} with the nature of the quantum
dynamics.  In order to proceed down that road we need to state
precisely a partition function based on this action, and
furthermore to construct the Feynman diagrams we require Feynman
rules.  We specify these explicitly in the form of propagators and
vertex operators which we extract from the kinetic and interaction
terms in the action, a standard modus operandi in field theory.
The split occurs as follows:
\begin{equation}\begin{split} S[\p,\psi_s]=&\;\frac{1}{2}\int
\prod^{3}_{i=1}dg_i\,d\bar{g}_i\;\p(g_1,g_2,g_3)\p(\bar{g}_1,\bar{g}_2,\bar{g}_3)\cK_{\p}(g_i,\bar{g}_i)\\
&+\int\prod^{6}_{i=1}dg_i\,d\bar{g}_i
\;\p(g_1,g_2,g_3)\p(\bar{g}_3,g_5,g_4)\p(\bar{g}_4,\bar{g}_2,g_6)\p(\bar{g}_6,\bar{g}_5,\bar{g}_1)\cV_{\p}(g_i,\bar{g}_i)\\
&+\frac{1}{2}\int dg_i\,d\bar{g}_i\,du\,d\bar{u}\;
\psi_{s}(g_1,g_2,g_3;u)\psi_{s}(g_1,g_2,g_3;u)\cK^{s}_{\psi}(g_i,\bar{g}_i,u,\bar{u})\\
&+\int \prod^{6}_{i=1}dg_i\,d\bar{g}_i\, du_a\, du_b\;
\psi_{s}(g_1,g_2,g_3;u_a)\psi_{s}(g_4,\bar{g}_3,g_5;u_b)\p(\bar{g}_4,\bar{g}_2,g_6)\p(\bar{g}_6,\bar{g}_5,\bar{g}_1)
\cV^{s}_{2\psi}(g_i,\bar{g}_i,u_a,u_b)\\ &+\int
\prod^{6}_{i=1}dg_i\,d\bar{g}_i\, du_a\, du_b\, du_c\;
\psi_{s}(g_1,g_2,g_3;u_a)\psi_{s}(g_4,\bar{g}_3,g_5;u_b)\psi_{s}(g_6,\bar{g}_4,\bar{g}_2;u_c)\p(\bar{g}_6,\bar{g}_5,\bar{g}_1)
\cV^{s}_{3\psi}(g_i,\bar{g}_i,u_a,u_b,u_c)\\ &+\int
\prod^{6}_{i=1}dg_i\,d\bar{g}_i\, du_a\, du_b\, du_c\, du_d\;
\psi_{s}(g_1,g_2,g_3;u_a)\psi_{s}(g_4,\bar{g}_3,g_5;u_b)\psi_{s}(\bar{g}_2,g_6,\bar{g}_4;u_c)\\
&\phantom{xxxxxxxxxxxxxxxx}\times\psi_{s}(\bar{g}_6,\bar{g}_5,\bar{g}_1;u_d)\cV^{s}_{4\psi}(g_i,\bar{g}_i,u_a,u_b,u_c,u_d).
\label{actionexp}\end{split}
\end{equation} The operators above are the kinetic and interaction operators
for the $\phi$ and $\psi$ fields, stated explicitly as: \beq
\cP_{\p}&\equiv&\cK_{\p}^{-1} = \cK_{\p}=\int d\a\,
d\bar{\a}\;\dk{1}\dk{2}\dk{3},\label{kp}\\
\cV_{\p}&=&\frac{\l}{4!}\int\prod_{i=1}^{4}d\a_i\;
\dvp{1}{4}{1}\dvp{2}{3}{1}\dvp{3}{2}{1}\dvp{4}{3}{2}\nonumber\\ &
&\hphantom{xxxxxxxxxxx}\times\dvp{5}{4}{2}\dvp{6}{4}{3},\label{vp}\\
\cP^{s}_{\psi}&\equiv&\cK_{\psi}^{s\;\;-1}=\cK^{s}_{\psi}=\int
d\a\, d\bar{\a}\;\dk{1}\dk{2}\dk{3}\d(\bar{u}\bar{\a}^{-1}\a
u^{-1}),\label{kc}\\
\cV^{s}_{2\psi}&=&\mu_2\int\prod_{i=1}^{4}d\a_i\;
\dvp{1}{4}{1}\dvp{2}{3}{1}\dvc{3}{2}{1}{a}\dvc{4}{3}{2}{b}\nonumber\\
&
&\hphantom{xxxxxxxxxxx}\times\dvp{5}{4}{2}\dvp{6}{4}{3}\d(u_a\a_1^{-1}\a_2u_b^{-1})\label{vc2}\\
\cV^{s}_{3\psi}&=&\mu_3\int\prod_{i=1}^{4}d\a_i\;
\dvp{1}{4}{1}\dvp{2}{3}{1}\dvc{3}{2}{1}{a}\dvc{4}{3}{2}{b}\nonumber\\
&
&\phantom{xxxxxxxxxxxx}\times\dvp{5}{4}{2}\dvp{6}{4}{3}\d(\a_1u_a^{-1}hu_a\a_1^{-1}\a_2u_b^{-1}h^{-1}u_b\a_2^{-1}\a_3u_c^{-1}hu_c\a_3^{-1})\nonumber\\
&
&\phantom{xxxxxxxxxxxxxxxxxxxxxx}\times\sum_{\substack{I_a,I_b,I_c\\
n_a,n_b,n_c}}D^{I_a}_{sn_a}(u_a\a_1^{-1})D^{I_b}_{sn_b}(u_b\a_2^{-1})D^{I_c}_{sn_c}(u_c\a_3^{-1})C^{I_a\,I_b\,I_c}_{n_an_bn_c},\label{vc3}\\
\cV^{s}_{4\psi}&=&\mu_4\int\prod_{i=1}^{4}d\a_i\;\dvp{1}{4}{1}\dvc{2}{3}{1}{d}\dvc{3}{2}{1}{a}\dvc{4}{3}{2}{b}\nonumber\\
&
&\phantom{xxxxxxxx}\times\dvp{5}{4}{2}\dvp{6}{4}{3}\d(\a_1u_a^{-1}hu_a\a_1^{-1}\a_2u_b^{-1}h^{-1}u_b\a_2^{-1}\a_3u_ch^{-1}u_c\a_3^{-1}\a_4u_d^{-1}hu_d\a_4^{-1})\nonumber\\
&
&\phantom{xxxxxxxxxxxxxxxx}\times\sum_{\substack{I_a,I_b,I_c,I_d,\L\\
n_a,n_b,n_c,n_d}}D^{I_a}_{sn_a}(u_a\a_1^{-1})D^{I_b}_{sn_b}(u_b\a_2^{-1})D^{I_c}_{sn_c}(u_c\a_3^{-1})D^{I_d}_{sn_d}(u_d\a_4^{-1})\tilde{C}^{I_a\,I_b\,I_c\,I_d\,\L}_{n_an_bn_cn_d},\label{vc4}
\eeq  where $\cP_{\p}$ and $\cP^{s}_{\psi}$ are the propagators
for the field theory.

Before we describe in detail the type of
amplitudes that arise once we implement the Feynman rules, let us
expound the attributes of the individual terms.

\begin{figure}[h]
\begin{center}
\includegraphics[width=13cm]{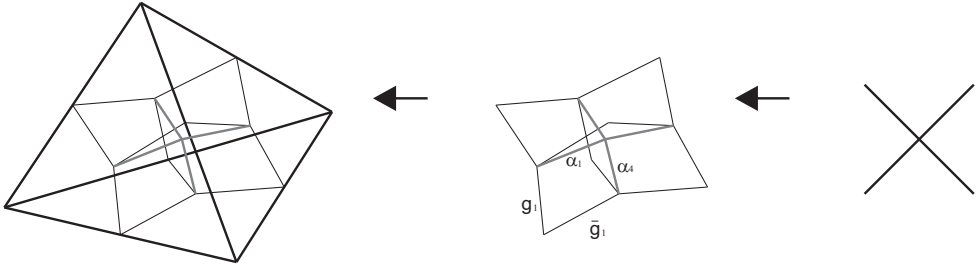}
\end{center}
\caption{\label{0tet}The pure gravity (Boulatov) tetrahedron.}
\end{figure}

The first two terms $S_{gr}$, with operators $\cP_{\p}$ and
$\cV_{\p}$, are those from Boulatov's group field theory. Thus,
pure gravity diagrams occur as a subset of graphs in our model.
Although the Boulatov vertex operator is well known, we describe
it in more detail here as later vertex operators are but
augmentations of this more basic structure.  The vertex term has
four $\p$ fields, thus four triangles, and the matching of their
arguments within is such that the four triangles they represent
form a tetrahedron. Moving on to the operator (\ref{vp}) itself,
we see that it contains six $\d$-functions.  Their arguments
represent holonomies around wedges dual to the edges of the
tetrahedron. We can see this diagrammatic structure in FIG.
\ref{0tet}.  The $\d$-functions force the holonomies to be the
identity which is the discrete analogue of forcing the wedge to
have zero curvature. We have a flat tetrahedron.

The propagator $\cP_{\p}$ for the pure gravity sector represents
geometrically the gluing of two tetrahedra obtained by identifying
one triangle from each tetrahedron. The third term is the kinetic
term for the coupled field. Its operator produces the propagator,
$\cP^{s}_{\psi}$, by inversion. The propagator is the identity on
function space. It has an important role to play in the
conservation of momentum as a particle travels from one
tetrahedron to the next and we will discuss this in more detail in
Section \ref{qdyn}, when we come to deal with generic Feynman
diagrams generated by the perturbative expansion of the partition
function.

\begin{figure}[h]
\begin{center}
\includegraphics[width=12cm]{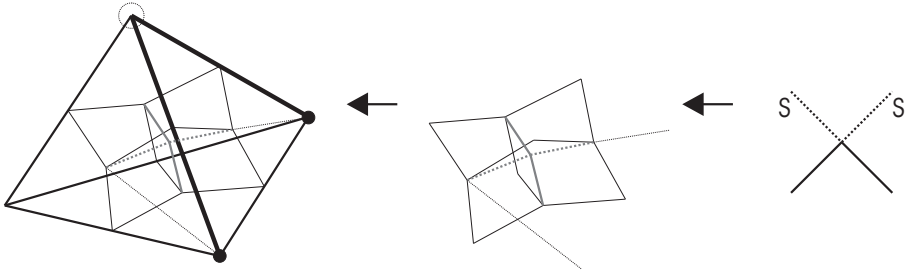}
\end{center}
\caption{\label{2tet}The tetrahedron with a bivalent particle
interaction.}
\end{figure}

The fourth term in the action $S_{gr+mat}^{int-2}$, with operator
$\cV^{s}_{2\psi}$, has two $\psi_s$ fields and two $\p$ fields.
This time two of the triangles have extra degrees of freedom
related to matter.  The arrangement of the gravity arguments once
again gives it the form of a tetrahedron.   In the operator
(\ref{vc2}), we have six $\d$-functions over the holonomies around
the wedges.  Only four are forced to be flat; two have defects
inserted.   We wrote the amplitude for the particle degrees of
freedom in a very simple form, however.  This hides a more
explicit description of the particle degrees of freedom and
furthermore, it does not look like a CPR spin foam building block.
We prove in Appendix \ref{simp} that the vertex term
$S_{gr+mat}^{int-2}$ satisfies the following equality:
\begin{equation}\begin{split}&\mu_2\int
\prod^{6}_{i=1}dg_i\, du_a\;
P_{\a_1}\psi_{s}(g_1,g_2,g_3u_a^{-1}hu_a;u_a)P_{\a_2}\psi_{s}(g_4u_b^{-1}h^{-1}u_b,g_3,g_5;u_a)
P_{\a_3}\p(g_4,g_2,g_6)P_{\a_4}\p(g_6,g_5,g_1)\\
&=\mu_2\int\prod^{6}_{i=1}dg_i\, du_a\, du_b\;
P_{\a_1}\psi_{s}(g_1,g_2,g_3u_a^{-1}hu_a;u_a)P_{\a_2}\psi_{s}(g_4u_b^{-1}h^{-1}u_b,g_3,g_5;u_b)
P_{\a_3}\p(g_4,g_2,g_6)P_{\a_4}\p(g_6,g_5,g_1)\\
&\phantom{xxxxxxxxxxxxxxxxxxxxxxxxx}\times
\d(u_a^{-1}hu_au_b^{-1}h^{-1}u_b)\sum_{n_a,n_b}D^{I}_{sn_a}(u_a)D^{I}_{sn_b}(u_b)\d_{n_an_b}.\label{biveq}\end{split}\end{equation}
The amplitude coming from this vertex term lends itself to a much
more explicit description of the particles' degrees of freedom.
\begin{equation}\begin{split}
\cV^{s}_{2\psi}=&\mu_2\int\prod_{i=1}^{4}d\a_i\;
\dvp{1}{4}{1}\dvp{2}{3}{1}\dvc{3}{2}{1}{a}\dvc{4}{3}{2}{b}\\
&\hphantom{xxxxxxxxxxx}\times\dvp{5}{4}{2}\dvp{6}{4}{3}\d(\a_1u_a^{-1}hu_a\a_1^{-1}\a_2u_b^{-1}h^{-1}u_b\a_2^{-1})\\
&\hphantom{xxxxxxxxxxxxxxxxxxx}\times\sum_{n_a,n_b}D^{I}_{sn_a}(u_a)D^{I}_{sn_b}(u_b)\d_{n_an_b},\label{vc2alt}
\end{split}\end{equation}
where $I$ is any fixed representation of $\SU(2)$ such that
$I-s\in\N$. We reiterate here that we label both particles, $a$
and $b$, by the same angular momentum $I$ and we do not sum over
$I$. This stems from the fact that a bivalent particle interaction
does not have any dependence on the total angular momenta and
summing redundantly over them would result in infinities. In fact,
we chose $I=s$ from now on. The defects we mentioned earlier are
the momenta $u^{-1}hu$ of the particles associated to two edges of
the tetrahedron. We illustrate them by the emboldened lines in
FIG. \ref{2tet}, and denote them by $\daleth$, the particle graph.
We denote the $\d$-function,
$\d(\a_1u_a^{-1}hu_a\a_1^{-1}\a_2u_b^{-1}h^{-1}u_b\a_2^{-1})$
imposing explicit momentum conservation by a dotted curve
encircling the vertex of the tetrahedron at which the two
particles interact.  This deals with the mass side of the
particles' degrees of freedom.  But we must also account for their
spin. The angular momenta, both the total and spin, reside on a
dual particle graph $\daleth^*$.  We draw this as the dashed line
in FIG. \ref{2tet}.  We label $\daleth^*$ by the matrix elements
of the holonomy along the dashed line in the total angular
momentum representation $I=s$.  At the endpoints of the
holonomies, we project the momenta of the particles down to the
spin-s
component:\begin{equation}D^{s}_{sn_a}(u_a\a_1^{-1})D^{s}_{sn_b}(u_b\a_2^{-1})\end{equation}
The bivalent interaction in $\daleth^*$ occurs at the dual vertex,
where we place the intertwiner $\d_{n_an_b}$.  We emphasise that
the topological equivalence of the $\daleth$ and $\daleth^*$ is an
imperative quality and determines in a large part how we define
our model (see Appendix \ref{cpr}).

\begin{figure}[h]
\begin{center}
\includegraphics[width=10cm]{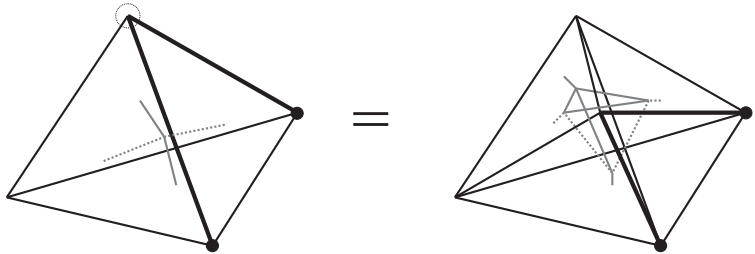}
\end{center}
\caption{\label{2teteq}The 1-4 equality satisfied by the bivalent
particle interaction term.}
\end{figure}

The vertex amplitude (\ref{vc2alt}), as we have defined it is
still not recognisable as a suitable building block for the
amplitudes of the CPR model. This is due to the presence of the
$\d$-function enforcing explicit momentum conservation, and indeed
the removal of this factor would supply us with a correct
amplitude for one tetrahedron with two particles present. As a
matter of fact the CPR spin foam amplitudes contain a type of
implicit momentum conservation, which we explain precisely in
Section \ref{ampdisc}. Importantly for us, the presence of this
$\d$-function means that our amplitude satisfies an equation given
pictorially in FIG. \ref{2teteq}.

This graphically represents the following equation:
\begin{equation}\begin{split}&\int\prod^{6}_{i=1}dg_i\,d\bar{g}_i\, du_a\, du_b\;
\psi_{s}(g_1,g_2,g_3;u_a)\psi_{s}(g_4,\bar{g}_3,g_5;u_b)\p(\bar{g}_4,\bar{g}_2,g_6)\p(\bar{g}_6,\bar{g}_5,\bar{g}_1)\\
&\hphantom{xxxx}\times\mu_2\int\prod_{i=1}^{4}d\a_i\;
\dvp{1}{4}{1}\dvp{2}{3}{1}\dvc{3}{2}{1}{a}\\
&\hphantom{xxxxxxxxxxxxxxxxxxxx}\times\dvc{4}{3}{2}{b}\dvp{5}{4}{2}\dvp{6}{4}{3}\\
&\hphantom{xxxxxxxxxxxxxxxxxxxxxxxxxxxxx}\times\d(\a_1u_a^{-1}hu_a\a_1^{-1}\a_2u_b^{-1}h^{-1}u_b\a_2^{-1})
\sum_{n_a,n_b}D^{I}_{sn_a}(u_a\a_1^{-1})D^{I}_{sn_b}(u_b\a_2^{-1})\d_{n_an_b}\\
&=\int\prod^{6}_{i=1}dg_i\,d\bar{g}_i\, du_a\,
du_b\int\prod_{i=1}^{4}d\a_i\prod_{j,k =1\; :\; j<k}^{4}
d\g_{jk}\;
\psi_{s}(g_1,g_2,g_3;u_a\g_{14}\a_1)\psi_{s}(g_4,\bar{g}_3,g_5;u_b\g_{24}\a_2)\p(\bar{g}_4,\bar{g}_2,g_6)\p(\bar{g}_6,\bar{g}_5,\bar{g}_1)\\
&\hphantom{xxxx}\times\mu_2
\dvg{1}{4}{1}\dvg{2}{3}{1}\dvg{3}{2}{1}\dvg{4}{3}{2}\dvg{5}{4}{2}\dvg{6}{4}{3}\\
&\hphantom{xxxxxxxxxxxxxx}\times\d(\g_{12}^{-1}\g_{24}^{-1}u_a^{-1}h^{-1}u_a\g_{14})\d(\g_{13}^{-1}\g_{23}\g_{12})\d(\g_{23}^{-1}\g_{34}^{-1}u_b^{-1}hu_b\g_{24})\d(\g_{13}\g_{14}^{-1}\g_{34})\\
&\hphantom{xxxxxxxxxxxxxxxxxxxxx}\sum_{n_a,n_b}D^{I}_{sn_a}(u_a)D^{I}_{sn_b}(u_b)\d_{n_an_b}.\label{2teteqmath}
\end{split}\end{equation}
where the $\g$ variables refer to the holonomies along the newly
introduced dual edges.  In words, the left hand side is the
amplitude $\cV_{2\psi}$ as it occurs in (\ref{vc2}). On the right
hand side the tetrahedron has been replaced by four tetrahedra;
this is the analogue of the 1-4 Pachner move of pure gravity in
the case in which particles are present. The particle graph
$\daleth$ has been \lq dragged' into the interior and does not
propagate along the original edges. Also we do not have explicit
momentum conservation at the vertex of $\daleth$.  So we see that
our vertex operator is a building block for a CPR amplitude with a
more refined triangulation.  We prove this equality explicitly in
Appendix \ref{onefour}.

\begin{figure}[h]
\begin{center}
\includegraphics[width=10cm]{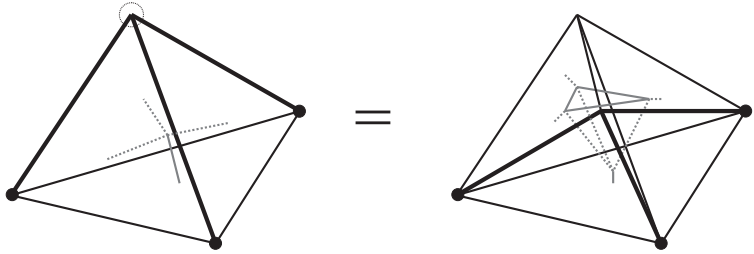}
\end{center}
\caption{\label{3teteq}The equality satisfied by the trivalent
particle interaction term.}
\end{figure}

The fifth term $S_{gr+mat}^{int-3}$, with operator
$\cV^{s}_{3\psi}$, has a slightly different structure to
$\cV^{s}_{2\psi}$, except there are three triangles with extra
degrees of freedom.  As usual, it has a tetrahedral structure, but
now three of the $\d$-functions have defects indicative of
particles on their associated edges. These form the particle graph
$\daleth$. Moreover, there is explicit momentum conservation.  For
the angular momenta, the structure of the operator identifies a
dual particle graph $\daleth^*$ with three edges and a trivalent
intertwiner at the dual vertex.  We mention also that the total
angular momenta of the particles are different as a trivalent
interaction depends on their individual values.  We sum to get the
most general amplitudes.  The two graphs, $\daleth$ and
$\daleth^*$, are again topologically equivalent.  Finally, the
explicit momentum conservation makes it again different at first
sight from the CPR amplitude for three particles on one
tetrahedron.  However, the very same $\d$-function is crucial for
showing the equality of this amplitude to an amplitude of the
CPR-type. This equality has the same form as that given in
(\ref{2teteqmath}) but with an extra particle added.  We represent
it pictorially in FIG. \ref{3teteq}.

\begin{figure}[h]
\begin{center}
\includegraphics[width=4cm]{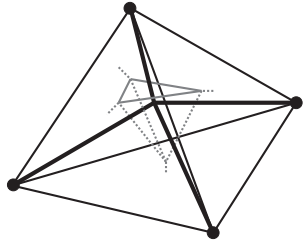}
\end{center}
\caption{\label{4tet}The definition of the 4-valent particle
interaction term.}
\end{figure}

The final term $S_{gr+mat}^{int-4}$, is that of the 4-valent
particle interaction. Every vertex of a tetrahedron is trivalent,
and therefore, to define a 4-valent particle interaction we must
do it in a situation where there is a 4-valent vertex at least.
Thus, it is clear that the vertex amplitude as we wrote it in
(\ref{vc4}) is hiding some information, that is, we have
integrated out some variables to write it in a simpler form.  But
as in all these cases the vertex term $S_{gf+mat}^{int-4}$
satisfies an equality:
\begin{equation}\begin{split}
&\mu_4\int\prod^{6}_{i=1}dg_i\,  du_a\, du_b\, du_c\, du_d\;
\psi_{s}(g_1,g_2,g_3u_a^{-1}hu_a;u_a)\psi_{s}(g_4u_b^{-1}h^{-1}u_b,g_3,g_5;u_b)\psi_{s}(g_2u_d^{-1}hu_d,g_6,g_4;u_c)\\
&\phantom{xxxxxxxxxx}\times\psi_{s}(g_6,g_4,g_2;u_d)\d(u_a^{-1}hu_au_b^{-1}h^{-1}u_bu_ch^{-1}u_cu_d^{-1}hu_d)\\
&\phantom{xxxxxxxxxx}\times\sum_{\substack{I_a,I_b,I_c,I_d,\L\\
n_a,n_b,n_c,n_d}}
D^{I_a}_{sn_a}(u_a)D^{I_b}_{sn_b}(u_b)D^{I_c}_{sn_c}(u_c)D^{I_d}_{sn_d}(u_d)\tilde{C}^{I_a\,I_b\,I_c\,I_d\,\L}_{n_an_bn_cn_d}\\
&=\mu_4\int\prod^{6}_{i=1}dg_i\,d\bar{g}_i \prod^{4}_{j=1}d\a_j\,
du_a\, du_b\, du_c\, du_d\;
\psi_{s}(g_1,g_2,g_3;u_a\g_{14}\a_1)\psi_{s}(g_4,\bar{g}_3,g_5;u_b\g_{24}\a_2)\psi_{s}(\bar{g}_2,g_6,\bar{g}_4;u_c\a_3)\\
&\phantom{xxxxxxxxxx}\times\psi_{s}(\bar{g}_6,\bar{g}_4,\bar{g}_2;u_d\a_4)\dvg{1}{4}{1}\dvg{2}{3}{1}\dvg{3}{2}{1}\\
&\phantom{xxxxxxxxxx}\times\dvg{5}{4}{2}\dvg{4}{3}{2}\dvg{6}{4}{3}\\
&\phantom{xxxxxxxxxx}\times
\d(\g_{12}^{-1}\g_{24}^{-1}u_a^{-1}h^{-1}u_a\g_{14})\d(\g_{13}^{-1}u_c^{-1}hu_c\g_{23}\g_{12})\d(\g_{23}^{-1}\g_{34}^{-1}u_b^{-1}hu_b\g_{24})\d(\g_{13}\g_{14}^{-1}u_d^{-1}h^{-1}u_d\g_{34})\\
&\phantom{xxxxxxxxxx}\times\sum_{\substack{I_a,I_b,I_c,I_d,\L\\
n_a,n_b,n_c,n_d}}
D^{I_a}_{sn_a}(u_a)D^{I_b}_{sn_b}(u_b)D^{I_c}_{sn_c}(u_c\g_{34}^{-1})D^{I_d}_{sn_d}(u_d)\tilde{C}^{I_a\,I_b\,I_c\,I_d\,\L}_{n_an_bn_cn_d},
\end{split}\end{equation}
where we may prove this equality by following the same procedure
as in Appendix \ref{onefour}. We illustrate this new vertex term
in FIG. \ref{4tet}.

We see that the vertex common to the four tetrahedra is 4-valent
and so can play host to the interaction of the momenta. In the
dual, any of the four dual vertices (by symmetry) will serve to
facilitate the interaction of the angular momenta. We do not have
any explicit momentum conservation here,  and so we have a CPR
building block. All further analysis follows that of the trivalent
interaction term. For that reason we do not mention this term for
the rest of the paper, but we state it here for completeness.

To conclude, we exhausted the possibilities for vertex terms. For
example, although the vertex term $\cV_{\psi}$ with one $\psi$
field allows the propagation of momentum along one edge of the
tetrahedron, the spin degree of freedom has no such path in the
dual as this requires two $\psi$ fields at least.  Meanwhile, we
cannot have more than 4-valent particle interactions in our model,
as the dual to a triangulation is 4-valent, and we must preserve
topological equivalence.

\subsection{Quantum dynamics and Feynman amplitudes}
\label{qdyn} We examine, in this section, the partition function
and transition amplitudes.  These define the quantum dynamical
aspects of our theory.  The dynamics has two facets:  non-perturbative aspects and
perturbative aspects.  We do not investigate non-perturbative
features as these are not well-understood even in the case of pure
gravity (but see the important work of \cite{instantons}).  In this work we focus on the perturbative features of our quantum
theory.  The first object of interest, which will receive most
attention, is the partition function, defined in perturbative expansion as:
\begin{equation}\begin{split} Z=&\int\cD\psi_s\cD\p\;
e^{-S[\p,\psi_s]},\\
=&\sum_{\G}\frac{\l^{v_0[\G]}\m_2^{v_2[\G]}\m_3^{v_3[\G]}}{sym[\G]}Z[\G],\end{split}\end{equation}
We denote each term in the
expansion by a Feynman diagram $\G$, and each $\G$ has an
amplitude $Z[\G]$. Moreover, $sym[\G]$ is the symmetry factor of
the graph, $v_0[\G]$ is the number of Boulatov tetrahedra,
$v_2[\G]$ is the number of tetrahedra with a bivalent particle
interaction and $v_3[\G]$ is the number of tetrahedra with a
trivalent particle interaction.  We may construct these terms in
the summation with the aid of a graphical calculus which we have
been developing over the course of the paper. We have a 4-valent
graph with two types of line, full and dashed, and three types of
vertex: four full lines incident, two full and two dashed, and one
full and three dashed.  Then we label them as below:
\begin{eqnarray}
\includegraphics[width = 6cm]{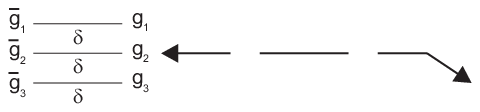} & & \cP_{\p}(g_i,\bar{g_i})\nonumber\\
\includegraphics[width = 6cm]{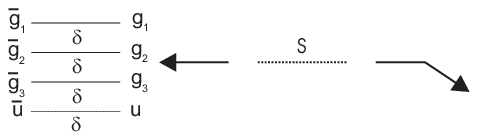} & &
\cP^{s}_{\psi}(g_i,\bar{g_i},u,\bar{u})\nonumber\\
\includegraphics[width = 6cm]{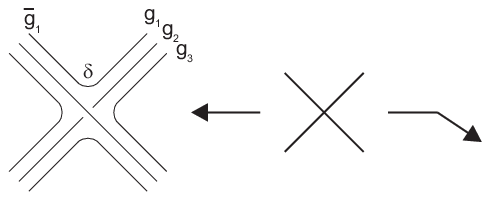} & &
\int\prod_{i=1}^{6}dg_i\,d\bar{g}_i\;\cV_{\p}(g_i,\bar{g_i})\nonumber\\
\includegraphics[width = 6cm]{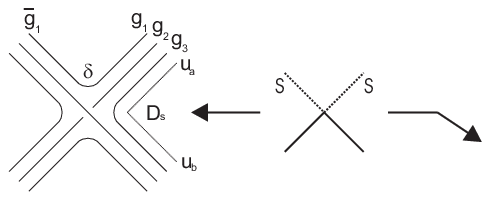} & &
\int\prod_{i=1}^{6}dg_i\,d\bar{g}_i\,du_a\,du_b\;\cV^{s}_{2\psi}(g_i,\bar{g_i},u_a,u_b)\nonumber\\
\includegraphics[width = 6cm]{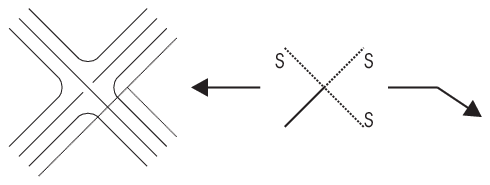} & &
\int\prod_{i=1}^{6}dg_i\,d\bar{g}_i\,du_a\,du_b\,du_c\;\cV^{s}_{3\psi}(g_i,\bar{g_i},u_a,u_b,u_c)\nonumber
\end{eqnarray}
Products of these form the Feynman graph and its structure is dual
to the structure of a triangulation. This is just conventional
Feynmanology for a path integral formulation of a field theory.

We can express the operators using strand diagrams.   They are an
intermediary calculus between the utter minimalism of the Feynman
graphs and the more complicated formalism of the fully
reconstruction of the triangulation.  To each pure gravity line we
associate three strands, and to each coupled line four strands. We
label the endpoints of the solid strands by the $g$ arguments, and
the endpoints of the dotted strand by the momentum of the particle
$u$.  We label the solid edges themselves by the $\d$-function
over the holonomy which contains the $g$ arguments of its two
endpoints, while the dotted strands are labelled by the angular
momentum amplitude.  Once vertices are glued using the
propagators, the solid strands form loops which are the plaquettes
dual to an edge of the triangulation and the dotted strands form
the dual particle graph $\daleth^*$.

We have yet to show that each term $Z[\G]$ is a spin foam
amplitude. A generic Feynman graph $\G$ in the partition function
is a closed 4-valent graph labelled as above. The integration over
$g$ and $u$ variables glues the vertex amplitudes, the
propagators being essentially identity operators. There are
basically three interesting subsets of diagrams that occur:
\begin{displaymath}\begin{array}{cccc}
v_0\neq0, &\quad\quad v_2=0, &\quad\quad v_3=0, &\quad\quad
\textrm{pure
gravity,}\qquad\qquad\qquad\qquad\qquad\qquad\quad\;\,\\ v_0\neq0,
&\quad\quad v_2\neq0, &\quad\quad v_3=0, &\quad\quad\textrm{non
interacting particles coupled to gravity,}\\ v_0\neq0, &\quad\quad
v_2\neq0, &\quad\quad v_3\neq0, &\quad\quad\textrm{interacting
particles coupled to gravity.}\quad\;\;\,
\end{array}\end{displaymath}
Consider the first scenario and the simplest. These are the
diagrams occurring in Boulatov field theory for pure gravity. Note
that each of the vertices of the Feynman graph is dual to a
tetrahedron and thus the integration of $g$ variables glues
tetrahedra together to form a triangulation.  To understand the
amplitudes we move to the dual picture where the aforementioned
integration glues wedges together to form faces each dual to an
edge of the triangulation.  Therefore, we are left with a
$\d$-function for the holonomy around each face of the dual which
enforces the flatness condition on the curvature.  This is exactly
the Ponzano-Regge spin foam amplitude for pure 3d Riemannian
quantum gravity \cite{laurentPRI}.

\begin{figure}[h]
\begin{center}
\includegraphics[width=17cm]{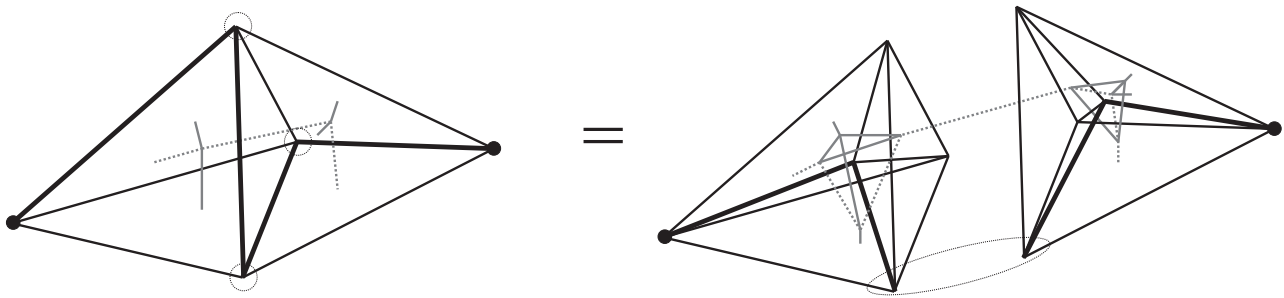}
\end{center}
\caption{\label{glu}The gluing of two coupled tetrahedra.}
\end{figure}

We now approach the more delicate task of constructing diagrams
including matter in their amplitudes.  The form of the amplitudes
can be extrapolated from the gluing of two vertices with bivalent
particle interactions. The contruction works analogously in the
case of 3-valent and 4-valent particle interactions, so we do not
discuss these other cases in detail. The propagator glues the
tetrahedra together at a face and ensures the conservation of
momentum explicitly at the vertex of the triangulation where the
segments of the particle graphs meet.  It also ensures the
conservation of spin in the dual particles graphs explicitly.  We
give this visually in FIG. \ref{glu}. 

\begin{figure}[h]
\begin{center}
\includegraphics[width=8cm]{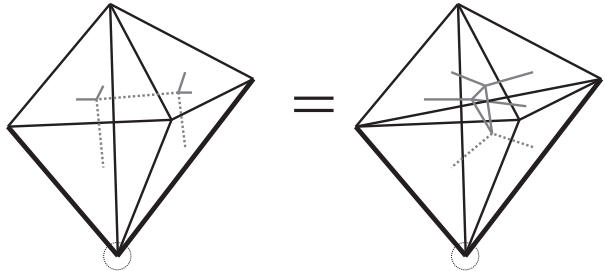}
\end{center}
\caption{\label{twothree}The 2-3 equality satisfied by the coupled
tetrahedra.}
\end{figure}

This can be written as a CPR amplitude very easily after some
rearrangement.  First we use the 1-4 equality to remove the signs
of explicit momentum conservation at two of the vertices (see
Appendix \ref{onefour}).  This is given in FIG. \ref{glu} also. We
remove the final explicit $\d$-function in two steps.  The
original shared face of the triangulation is now shared by two
smaller tetrahedra.  These are drawn on the lhs of FIG.
\ref{twothree}.  Each of the tetrahedra has a particle on an edge.
We know we can use the 1-4 equality to remove explicit momentum
conservation when the particles are in the same tetrahedron.
Fortunately, the 2-3 Pachner equality is satisfied by these two
tetrahedra when particles are present.  We prove this in Appendix
\ref{twothreesect} and draw it here in FIG. \ref{twothree}.  Now
that the two particles are in the same tetrahedron, a final 1-4
equality can be invoked. We arrive at an amplitude that has no
explicit signs of momentum conservation and is a CPR amplitude.

In effect the amplitude we see on gluing two of our vertices
together is the coarse graining of a CPR spin foam amplitude for a
finer triangulation, i.e. with some of the arguments integrated
out. The above reorganisation  is indicative of the procedure when
two coupled tetrahedra are glued together, and gives us the
insight necessary to chronicle the form of a generic Feynman graph
for matter coupled to gravity.  For a normal edge of the
triangulation we get the usual curvature flatness constraint.  For
an edge of the particle graph $\daleth$, we find that there is a
defect in the curvature equal to the momentum of the particle. For
the dual particle graph $\daleth^*$, we labelled it by the matrix
elements of its holonomy in the total angular momentum
interspersed with spin projections, one for each particle, and
angular momentum intertwiners one for each particle interaction.
This is exactly a CPR spin foam amplitude. 

\begin{figure}[h]
\begin{center}
\includegraphics[width=5cm]{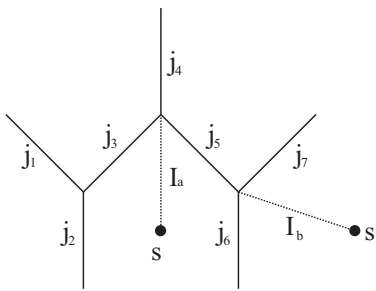}
\end{center}
\caption{\label{bound}Portion of a typical boundary state.}
\end{figure}

The next object is a definition of the transition amplitudes.
fortunately, we have done the background work on this already so
we can proceed and state it as the \lq two point' or more
precisely, \lq two net' function:
\begin{equation}<\o_1|\o_2>=\int\cD\psi_s\cD\p\;
\o_1[\p,\psi_s]\o_2[\p,\psi_s]e^{-S[\p,\psi_s]}\end{equation}
where $\o_1$ and $\o_2$ are the boundary states, i.e. products of
the Boulatov and coupled fields based on open spin networks.  We
give a portion of a typical boundary state in FIG. \ref{bound}.
FIG. \ref{bound} evaluates to
\begin{equation}\dots\P^{j_1\;\;j_2\;\;j_3}_{m_1m_2m_3}\Psi^{j_3\;\;j_4\;\;j_5\;\,I_a\;\L_a}_{m_3m_4m_5s}\Psi^{j_6\;\;j_5\;\;j_7\;\,I_b\;\L_b}_{m_6m_5m_7s}\dots\end{equation}
with repeated indices summed over.  This concludes our definition
of the model.

\section{Features of the new model - extended discussion}
\subsection{Feynman Amplitudes}
\label{ampdisc} In Section \ref{model} we detailed the definition
of our model but we postponed almost all explanation of our
reasons for the particular form we chose. The model in question
has many intriguing characteristics which deserve further
clarification since they play non-obvious roles in ensuring the
faultless realisation of the Feynman amplitudes as CPR spin foam
amplitudes. Among these, two are most notable: the amplitudes are defined with explicit momentum
conservation at every vertex of $\daleth$; also, we do not impose
permutation invariance in the coupled field.

Before we elucidate our reasoning we give some background
information on the CPR amplitudes that is especially relevant
here.  The topic is the implicit momentum conservation contained
in the CPR spin foam model.  For sake of completeness, we start
from the very beginning.  Classically, first order 3d Riemannian
gravity is a BF theory.  The curvature $F$ satisfies a Bianchi
identity, $dF=0$, and as $dF$ is a 3-form it couples to volumes.
Upon discretising the manifold and quantising the theory, the
continuum Bianchi identity is lost, understandably, but a discrete
remnant survives. We describe this by considering a vertex of the
triangulation $\D$, and all the edges emanating from it.  Each of
these edges is dual to a face of the spin foam and therefore has a
holonomy associated to it. These faces close to form a 3-ball
around the vertex and thus identify the boundary of a volume. The
discrete version of the Bianchi identity states that there exists
an ordered product of these holonomies which evaluates to the unit
element \cite{laurentPRI}. For pure gravity, since the very
quantum amplitude enforces the flatness of the holonomy this
identity gives us no more restrictions on the dynamics of the
theory\footnote{In fact it causes an infinity to appear in the
amplitude, which signals that their is a gauge symmetry which
needs to be fixed \cite{laurentdiffeo}.}. Now, consider a vertex
where $n$ of the edges have particles propagating along them.  The
amplitude for these edges is a $\d$-function over the holonomy
with a momentum defect. Thus, the holonomy is forced to be the
momentum of the particle and the product of holonomies in the
Bianchi Identity reduces to a product over the momenta incident at
the vertex. Thus, we have implicit {\it overall} momentum
conservation. This is not sufficient for our purposes, however.
Unlike the spin foam quantisation, where one has an explicit
knowledge of the discretisation structure, i.e. the triangulation
$\D$ and the dual spin foam $\D^*$, before one begins, and one has
hands-on control over the particle graphs, $\daleth$ and
$\daleth^*$ and how one labels them, i.e. what are the variables
living on them, the group field theory does not allow such freedom
and control. We generate the spin foams, after all, as Feynman
graphs and we only know the action to start with.  To illustrate
the way momentum conservation is realised in our model and the
topological equivalence between $\daleth$ and $\daleth^*$ is
mantained, in spite of this lack of control, we develop two
examples.

\begin{figure}[h]
\begin{center}
\includegraphics[width=10cm]{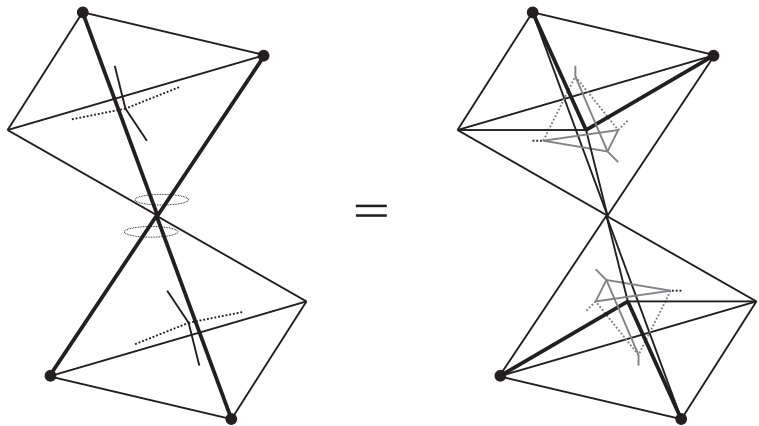}
\end{center}
\caption{\label{multivert}Two tetrahedra sharing the same vertex
(both having a bivalent interaction there).}
\end{figure}

As our first case, since in our Feynman expansion we generate all
possible graphs given the Feynman rules, one possibility is that
in FIG. \ref{multivert}.  These tetrahedra have one vertex in
common and it happens to be the one where both of them have a
bivalent particle interaction in $\daleth$. Implicit momentum
conservation here would imply that all the four momenta incident
to the vertex would sum up to zero, with orientation taken into
account, but this would allow for the identification of a 4-valent
matter interaction vertex on $\daleth$, instead of two bivalent
ones.  This disagrees with the two bivalent particle interactions
in $\daleth^*$, and so we would not have a CPR amplitude, due to
the breaking of the equivalence of $\daleth$ and $\daleth^*$. In
other words, the implicit momentum conservation that we know is
present in the Feynman amplitudes is not enough to guarantee the
well-posedness of the model. Since we have explicit momentum
conservation, this scenario does not arise and we have the correct
particle graph structure. It is gratifying also to see that the
introduction of explicit momentum conservation means that we can
separate the two particle interactions by the \lq dragging'
procedure of the 1-4 equality. In this way the ambiguous situation
discussed above is simply removed, as $\daleth$ and $\daleth^*$
have now manifestly the same structure. This confirms the more
fundamental nature of the dual particle graph $\daleth*$ (at least
in our model) and that ambiguous configurations arise only as a
result of \lq pathological' embeddings of the dual particle graph
in the triangulation, or equivalently, as a result of coarse
graining of the same triangulation, performed by means of Pachner
moves.

The second example displays a different problem and its solution.
Another possibility in a Feynman diagram is that we generate two
tetrahedra that share an edge which happens to be an edge where
both tetrahedra have a particle propagating. This could be at
first sight interpreted indeed as the presence of two particles
propagating along the same edge of the triangulation, and the
implicit momentum conservation would be compatible with this
interpretation. This type of configuration does not occur in the
CPR amplitudes of \cite{laurentPRI}, and is again a pathological
result of inappropriate coarse graining of the triangulation.
Their pathological nature is clear if we recall that we interpret
the particle graphs as Feynman graphs of a matter field theory
coupled to gravity, so that lines should represent one-particle
propagation, and that the CPR amplitudes reduce indeed in the
effective limit to Feynman diagrams of a non-commutative quantum
field theory, so they lie within a larger multiparticle structure.
We neither need nor want multiple particles on an edge. With 1-4
equality, i.e. by application of a Pachner move on the
triangulation labelled by particle degrees of freedom, this
pathology can be overcome and the particle graph \lq resolved'
into a physically equivalent one (because it has equal amplitude)
that shows no ambiguous \lq multi-particle' appearance.

\begin{figure}[h]
\begin{center}
\includegraphics[width=15cm]{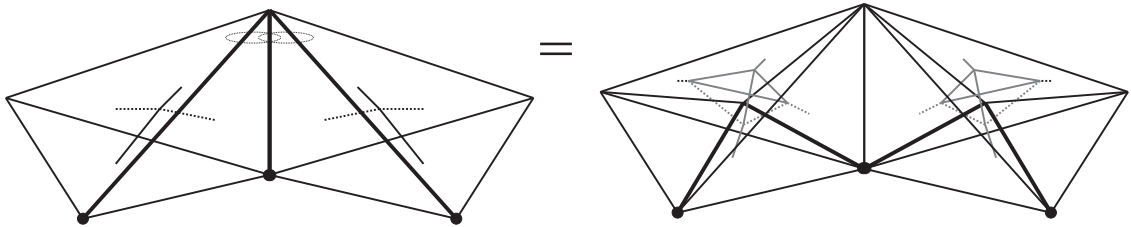}
\end{center}
\caption{\label{multiedge}Two tetrahedra sharing the same edge
(both contributing a particle there).}
\end{figure}

\begin{figure}[h]
\begin{center}
\includegraphics[width=7cm]{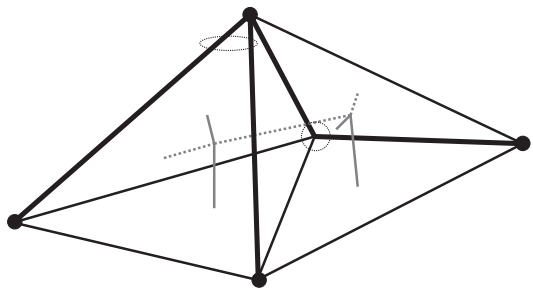}
\end{center}
\caption{\label{badperm}A configuration coming from requiring
permutation invariance.}
\end{figure}

Again, the crucial point is the topological equivalence of
$\daleth$ and $\daleth^*$, so let us discuss this a bit more. In
the spin foam context where we have complete control of the
variables it is natural to consider the particle graph residing in
$\D$ as more fundamental and the particle graph $\daleth^*$ in
$\D^*$ as a framing of $\daleth$ in the dual which gives a
consistent picture of how the angular momenta propagates.  This is
not, however, the way we picture things when dealing with the
amplitudes generated by the GFT, even though they are the same as
those of the CPR spin foams.  We perform a conceptual shift.  This
begins with the proposal of a new field to describe matter coupled
to gravity.  We see that this helps create the sense of the
particle graph $\daleth^*$ in $\D^*$ as our initial concept. In
other words, our Feynman expansion creates first a particle graph
$\daleth^*$ alongside the dual complex from which the
triangulation is reconstructed; from this, taking into account the
way the variables associated with this dual graph $\daleth*$ are
coupled with the gravity degrees of freedom, one has to
reconstruct a particle graph $\daleth$, lying on the triangulation
itself. In fact, the positioning of the momentum in the field
gives rise to the particle graph $\daleth$ when we reconstruct
$\D$ and all our efforts have been to ensure that $\daleth$ is
topologically equivalent to $\daleth^*$. Thus we can consider
$\daleth$ as an embedding of $\daleth^*$ into the triangulation.
It is then easy to see that topological equivalence forces us not
to impose the usual permutation invariance of pure gravity on the
coupled field.  To exemplify the type of problematic
configurations that arise were we to require permutation
invariance in the coupled field, consider the structure
obtained by gluing two tetrahedra as in FIG. \ref{badperm}. This is similar to the gluing
of two tetrahedra in FIG. \ref{glu}.  The edges on the shared face
of one tetrahedron, however, are permuted cyclically with respect
to the edges of the other.

The particle graph $\daleth$ no longer describes the continuous
propagation of a particle, even though momentum is conserved.
Topological equivalence is lost.

We should also discuss to what extent our model encompasses the
CPR spin foam amplitudes in all their generality. For spinning
particles, topological equivalence requires that the particle
graphs of the CPR spin foams have at most 4-valent interactions.
So, no generality is lost there.  However, if spinless particles
are dealt with as a special case of our more general formalism,
the same restriction would apply to them, while in the spin foam
amplitudes of the CPR model no limit on the order of interaction
for a scalar field was imposed.  Thus at first sight it seems that
some generality may be lost; it is not easy on the other hand to
confirm this impression nor to disprove it, given that we are able
to relate certain graphs arising in our model using Pachner moves,
and that this is possible also in the CPR amplitudes; by doing so
higher valent interaction vertices for scalar particles can be
resolved into lower valent ones, as well as the opposite.
Furthermore, we note that it is possible as well to generalise our
model to one that produces higher valent vertices even for
spinning particles/fields: we would have to generalise the
discretisation structures arising from the perturbative expansion,
and in particular those structures we reconstruct from Feynman
vertices from tetrahedra to polyhedra, as the dual vertex would
then have a higher number of incident dual edges. In any case we
see that a model with a restriction on interaction vertices to be
4-valent would thus look more fundamental in nature.

\subsection{Many particle species}

To this point, our model incorporated just one species of particle
with a fixed mass and spin, albeit arbitrary.  We extend our model
here to include many species of particle. This is done easily at
the dynamical level by adding new kinetic and vertex terms to the
action, with a similar structure to those already present. We
write down the terms in a new shorthand notation, that should,
however, be of no difficult to interpret. In this notation, our
original action is
\begin{equation}S[\p,\psi_{s,m}]=S_{\p\p}+S_{\p\p\p\p}+S_{\psi_{s,m}\psi_{s,m}}+S_{\psi_{s,m}\psi_{s,m}\p\p}+S_{\psi_{s,m}\psi_{s,m}\psi_{s,m}\p}.\end{equation}
The new terms are of such a similar form that we can extend as
follows:
\begin{equation}S[\p,\psi_{s,m},\psi_{\bar{s},\bar{m}}]=S[\p,\psi_{s,m}]+S_{\psi_{\bar{s},\bar{m}}\psi_{\bar{s},\bar{m}}}
+S_{\psi_{\bar{s},\bar{m}}\psi_{\bar{s},\bar{m}}\p\p}
+S_{\psi_{\bar{s},\bar{m}}\psi_{\bar{s},\bar{m}}\psi_{\bar{s},\bar{m}}\p}
+S_{\psi_{\bar{s},\bar{m}}\psi_{s,m}\psi_{s,m}\p}
+S_{\psi_{\bar{s},\bar{m}}\psi_{\bar{s},\bar{m}}\psi_{s,m}\p}
\end{equation}
The first three are just a replica of the terms given for the
$(s,m)$ particle earlier.  The final two allow for interaction
between the two species.  Further generalisations follow the same
path.

\begin{figure}[h]
\begin{center}
\includegraphics[width=4cm]{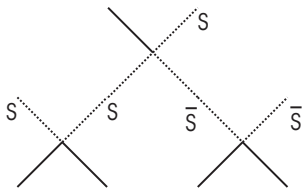}
\end{center}
\caption{\label{multiint}A generic multi-species Feynman diagram.}
\end{figure}

We do not specify the Feynman rules for the partition function and
transition amplitude explicitly but we describe the generic
structure of the diagrams that occur in the partition function,
and their particle graphs. Since the action contains the action of
\ref{action}, we get all the diagrams that we had before.
Furthermore, there is a subset of terms for the
$(\bar{s},\bar{m})$ particles coupled to gravity that are a direct
copy of the terms for $(s,m)$.  This means we get a copy of the
$(s,m)$ particle diagrams for the $(\bar{s},\bar{m})$ particles.
Finally, we get the diagrams charting the interaction between the
two species.  We give a portion of a typical example in FIG.
\ref{multiint}.

Kinematically, we can have boundary states containing particles of
both species, $\o(\phi,\psi_{s,m},\psi_{\bar{s},\bar{m}})$. These
have the same form as before.  They are based on open trivalent
graphs where some of the particle vertices are labelled by the
$\psi_{\bar{s},\bar{m}}$ and some by $\psi_{s,m}$.  Thus, there
exist non-zero transition amplitudes between multi-species,
multi-particle states.

\subsection{DSU(2) structures and Lorentz deformation}
Let us discuss briefly the role of non-commutative and
specifically $DSU(2)$ structures in our model. It is known that
quantum point particles in 3d manifest a symmetry under a quantum
group deformation of the Poincar\'{e} group, i.e. the non-compact
double of $SU(2)$, $DSU(2)$
\cite{matschullwelling,laurentPRI,laurentPRII,jurleelau}, and that
this deformed symmetry structure can be identified clearly, made
explicit and put to use in the CPR spin foam model
\cite{laurentPRI, laurentPRII} and in the effective field theory
for scalar matter fields derived from it \cite{laurentPRIII}.
Therefore, one may wonder what role $DSU(2)$ structures play in
our GFT model, and maybe expect that a correct group field theory
derivation of the CPR spin foam model should be based directly on
such $DSU(2)$ structures instead of using only the $SU(2)$ group
elements and representations as is the case for our model. We
think this is not the case and the evidence that can be gathered
from the literature suggests that, while a $DSU(2)$ symmetry is
likely to be implicitly present in our model, a purely
$SU(2)$-based formulation of 3d gravity coupled to matter at the
group field theory level is in our opinion not only sensible, but
arguably the most natural way to proceed.

The main reason for this is that not only the expression of the
CPR spin foam model, but also both its known derivations from
covariant path integral methods in a discrete setting
\cite{laurentPRI} and from canonical Hamiltonian methods in the
continuum \cite{KarimAlex}, make use of elements and
representations of $SU(2)$, and of its inhomogeneous counterpart
(i.e. the Poincar\'e group), only. In \cite{laurentPRI}, the
authors started with a discretisation of the continuum action for
3d gravity coupled to point particles, which is invariant under
local $SU(2)$ transformations only, and with the particles
labelled by Poincar\'{e} representations. They then derived, using
spin foam techniques, the CPR model where, as explained in detail
in Appendix \ref{cpr}, the modified amplitudes are functions of
$SU(2)$ group elements and group representations only. Also,
boundary data for the partition function are ordinary open spin
networks again based on $SU(2)$. Analogously, one can start from a
conventional definition of canonical kinematical states for matter
coupled to quantum gravity in terms of open $SU(2)$ spin networks
and Poincar\'{e} representations. Then, one can define
\cite{KarimAlex} and explicitly construct, using a rigorous
discretisation and application of usual loop quantum gravity
methods, a projection operator onto solutions of the Hamiltonian
constraint and a physical scalar product for canonical states
given again by the CPR spin foam model. The interesting point is
that in spite of this more conventional-looking expression of the
partition function, one can identify \cite{laurentPRI} a
non-trivial braiding for the particles. This results from the
action of the braiding matrix of the quantum group $DSU(2)$. Also,
the {\it same} partition function, for a given coupled Feynman
graph for matter fields, can be re-expressed \cite{laurentPRII} as
the evaluation of a colored chain mail link based again on the
same quantum group. These results make clear that quantum group
symmetries are a result of the {\it quantum
  dynamics} of gravity coupled to particles and not of the kinematics
behind it. But these results also make clear that the use of
un-deformed $SU(2)$ structures and of related spin foam techniques
for describing these quantum dynamics is fully compatible with the
presence of deformed symmetries for matter fields and with the use
of quantum group techniques for the evaluation of the same
physical quantities.

Let us also stress, in support of this conclusion, that the
equivalence of the Ponzano-Regge spin foam model, expressed and
derived {\it only} using $SU(2)$ structures, with a quantum group
evaluation of a chain mail link based on $DSU(2)$ is true also in
the simpler case of pure gravity with no matter coupling
\cite{laurentPRII}. As for the group field theory derivation of
such a model, it is well known that this is given by the Boulatov
group field theory \cite{boulatov}, i.e. by a field theory over an
ordinary $SU(2)$-based  group manifold, in spite of this
alternative reformulation based on $DSU(2)$.

There is more. The most striking appearance of non-commutative
structures from the CPR model is in our opinion the derivation of
an effective non-commutative field theory for a scalar field
encoding the quantum gravity corrections \cite{laurentPRIII}.
Here, it is clear that quantum gravity dynamics is responsible for
the deformation of ordinary field theory to a non-commutative one.
Again, however, in momentum space the resulting field theory is
just one based on an $SU(2)$ group manifold. The non-commutative
structure of spacetime emerges only after harmonic analysis, due
to the curvature of momentum space \cite{laurentPRIII}. The
$DSU(2)$ symmetry manifests itself not at the level of the action
but when considering multiparticle states, their non-trivial
braiding and their modified scattering laws. It is a field theory
of this type, based on the $SU(2)$ group manifold, that we expect
to obtain from a group field theory action, at an effective level,
after suitable integration of the gravity degrees of freedom.

In spite of all this evidence for the adequacy of using only
$SU(2)$ structures for constructing a group field theory
describing 3d quantum gravity coupled to matter fields, one may
still want to look for an alternative formulation that makes
explicit use of the quantum group $DSU(2)$; perhaps for the need
for greater simplicity or simply because of the beautiful
mathematical structures this would bring into play. This would
certainly be an interesting and fascinating project, but the
results of \cite{kirillgft} lead one to approach this issue with
greater caution. As already mentioned, in fact, in
\cite{kirillgft} the author studies the straightforward
generalisation of the Boulatov model to the case of $DSU(2)$, and
the whole construction proceeds beautifully and rigorously, but
the resulting model does not admit any clear interpretation in
terms of matter fields coupled to gravity (and definitely does not
reproduce the CPR spin foam model) that can be considered as a
sensible coupling of 3d quantum gravity with Feynman graphs for
matter fields. This suggests that something more elaborate may be
needed. We leave this for future work.

\subsection{Generalised model with variable mass and spin}

An interesting generalisation which has received some attention
\cite{kirillgft} (and we expect it to receive more shortly
\cite{karim}) is to relax our constraints on the mass and spin
completely.  To state things explicitly, we integrate over all
masses and sum over every spin. From one base, we could regard
this as allowing for every possible spin given a certain mass.  Or
from the other, for a given spin every mass is possible.  Thus
there is an infinite number of species of particles.  We give the
action as:
\begin{equation}\begin{split} S[\p,\psi]=&\;\frac{1}{2}\int
\prod^{3}_{i=1}dg_i\;P_{\a}\p(g_1,g_2,g_3)P_{\bar{\a}}\p(g_1,g_2,g_3)\\
&+\frac{\l}{4!}\int\prod^{6}_{i=1}dg_i\;P_{\a_1}\p(g_1,g_2,g_3)P_{\a_2}\p(g_3,g_5,g_4)P_{\a_3}\p(g_4,g_2,g_6)P_{\a_4}\p(g_6,g_5,g_1)\\
&+\frac{1}{2}\int dg_i du\;
\sum_{s}P_{\a}\psi_{s}(g_1,g_2,g_3;u)P_{\bar{\a}}\psi_{s}(g_1,g_2,g_3;u)\\
&+\mu_2\int \prod^{6}_{i=1}dg_i\, du_a \int dh \sum_{s}
P_{\a_1}\psi_{s}(g_1,g_2,g_3u_a^{-1}hu_a;u_a)P_{\a_2}\psi_{s}(g_4u_a^{-1}h^{-1}u_a,g_3,g_5;u_a)\\
&\phantom{xxxxxxxxxx}\times
P_{\a_3}\p(g_4,g_2,g_6)P_{\a_4}\p(g_6,g_5,g_1)\\ &+\mu_3\int
\prod^{6}_{i=1}dg_i\, du_a\, du_b\, du_c\; \int dh_a\,dh_b\,dh_c
\sum_{s_a,s_b,s_c}
P_{\a_1}\psi_{s_a}(g_1,g_2,g_3u_a^{-1}h_au_a;u_a)P_{\a_2}\psi_{s_b}(g_4u_b^{-1}h_b^{-1}u_b,g_3,g_5;u_b)\\
&\phantom{xxxxxxxxxx}\times
P_{\a_3}\psi_{s_c}(g_6,g_4,g_2u_c^{-1}h_cu_c;u_c)P_{\a_4}\p(g_6,g_5,g_1)\d(u_a^{-1}h_au_au_b^{-1}h_b^{-1}u_bu_c^{-1}h_cu_c)\\
&\phantom{xxxxxxxxxxxxxxxxxxxx}\times\sum_{\substack{I_a,I_b,I_c\\
n_a,n_b,n_c}}
D^{I_a}_{s_an_a}(u_a)D^{I_b}_{s_bn_b}(u_b)D^{I_c}_{s_cn_c}(u_c)C^{I_a\,I_b\,I_c}_{n_an_bn_c}
\label{genact}\end{split}\end{equation} We should note that in the
second vertex term, that is the vertex with a bivalent particle
interaction, we have written the two particles with the same mass
and spin because even if we allowed them to differ, the relations
above would force the amplitude to be zero except when they
coincided, due to momentum conservation. On the boundary, we
retain states with fixed mass and spin for any one particle, and
one can obviously have more than one species of particle in a
kinematic state. This is what we measure in practice, i.e. in real
life situations, and we wish to retain some sort of touch with
possible future experiment. This loosening of restrictions affects
the amplitudes, of course. More strikingly, it alters the
effective limit of the theory, and indeed the abelian limit of the
theory where one recovers usually ordinary quantum field theory.
This group field theory reduces to analogous theories but with the
sum over all masses and spins maintained, thus to field theories
with variable mass and spin; while we are aware of past work on
quantum field theories with indefinite mass (see for example
\cite{feynman, hostler}), we do not recall a similar
generalisation for the spin degrees of freedom.

\subsection{Reduced model - scalar fields}

Let us consider the limiting case in which both the spin of the
particle and its total angular momentum go to zero, i.e. the case
of 3d quantum gravity coupled to a single scalar field with no
angular momentum.  In \cite{us}, a model was proposed to describe
such happenstance, based on a formalism, a type of field and
relative action that are quite different from the ones we use in
the corresponding limit of our present model. However, they
produce, modulo multiplicative factors, the same Feynman
amplitudes. It is interesting then to understand the exact
relation between these two models. We converge on a reconciliation
first from the side of our new model.  The action for scalar
particles reduces to:
\begin{equation}\begin{split} S[\p,\psi]=&\;\frac{1}{2}\int
\prod^{3}_{i=1}dg_i\;P_{\a}\p(g_1,g_2,g_3)P_{\bar{\a}}\p(g_1,g_2,g_3)\\
&+\frac{\l}{4!}\int\prod^{6}_{i=1}dg_i\;P_{\a_1}\p(g_1,g_2,g_3)P_{\a_2}\p(g_3,g_5,g_4)P_{\a_3}\p(g_4,g_2,g_6)P_{\a_4}\p(g_6,g_5,g_1)\\
&+\frac{1}{2}\int dg_i du\;
P_{\a}\psi_{0}(g_1,g_2,g_3;u)P_{\bar{\a}}\psi_{0}(g_1,g_2,g_3;u)\\
&+\mu_2\int \prod^{6}_{i=1}dg_i\, du_a\;
P_{\a_1}\psi_{0}(g_1,g_2,g_3u_a^{-1}hu_a;u_a)P_{\a_2}\psi_{0}(g_4u_a^{-1}h^{-1}u_a,g_3,g_5;u_a)P_{\a_3}\p(g_4,g_2,g_6)P_{\a_4}\p(g_6,g_5,g_1)\\
&+\mu_3\int \prod^{6}_{i=1}dg_i\, du_a\, du_b\, du_c\;
P_{\a_1}\psi_{0}(g_1,g_2,g_3u_a^{-1}hu_a;u_a)P_{\a_2}\psi_{0}(g_4u_b^{-1}h^{-1}u_b,g_3,g_5;u_b)P_{\a_3}\psi_{0}(g_6,g_4,g_2u_c^{-1}hu_c;u_c)\\
&\phantom{xxxxxxxxxx}\times
P_{\a_4}\p(g_6,g_5,g_1)\d(u_a^{-1}hu_au_b^{-1}h^{-1}u_bu_c^{-1}hu_c)
\end{split}\end{equation}
We have no longer spin and angular momentum degrees of freedom
identifying a dual particle graph $\daleth^*$, we are only
interested in the particle graph $\daleth$, where we still have
explicit momentum conservation, and thus have at most trivalent
particle interactions (as we have not included 4-valent
interaction terms in the above action). No restriction comes then
from the need to have topological equivalence of $\daleth$ and
$\daleth^*$. The above action produces, in perturbative expansion,
the CPR amplitudes for scalar particles. The presence of explicit
momentum conservation, i.e. of extra deltas relating the $u$
variables, allows for the use of Pachner moves to resolve
pathological multiparticle-like configurations (in the sense
explained above). If we go one step further and remove explicit
momentum conservation, we recover a model in which multiple
particles reside on some edges and we have arbitrary valence
interactions in the particle graph $\daleth$. These are {\bf not}
CPR amplitudes, and correspond to the amplitudes obtained using
the alternative model presented in section IVC of \cite{us} in
which the extra variables labelling the field of the main model
have been removed. As explained in \cite{us} the extra structure
of the field (extra three arguments labelling possible particle
degrees of freedom) serve exactly the purpose of avoiding the
appearance of such multiparticle configurations, while retaining
the possibility of arbitrary valence of interaction and not
imposing momentum conservation explicitly.

Now let us approach this problem from the other side. In the model
of \cite{us} the number of arguments in the field is doubled and
the new ones are identified with the particle degrees of freedom
but in a manner different from what we are used to in this paper.
They are not used to ensure explicit momentum conservation at the
vertices and so we can have an arbitrary valence of interaction.
This is perfectly fine for scalar particles in the CPR model where
no topological equivalence is required.  Further, these new
variables are used to propagate information around a dual face so
that, should there be multiple particles on an single edge, they
cancel out in pairs, so that in the end one is left with either
one or zero particles on an edge.  The problem of multiparticle
configuration was thus solved in a very different way from the one
we adopted here.  The introduction of these new variables also had
the effect of increasing the order of infinity of the graphs above
that of the CPR amplitudes; this is attributed to redundant
additional gauge symmetry with respect to the pure gravity case,
which must be fixed by some procedure. Modulo these infinities,
the model indeed generates CPR amplitudes for scalar particles
coupled to quantum gravity.  As we said above, if one removes the
doubling of arguments in the field, one recovers a model,
presented as a possible alternative in \cite{us}, in which
multiple particles reside on some edges and we have arbitrary
valence interactions in the particle graph $\daleth$. However,
these are {\bf not} CPR amplitudes.

So we can see more clearly now, that the model proposed in
\cite{us} and the one we obtain here in the special case of scalar
fields, solve the problem of matter coupling to 3d quantum gravity
in two very different but easily related ways. The new model,
however, allows for the description of other types of fields as
well, and when non-zero spin and non-zero angular momentum are
considered, it has the added responsibility to ensure that the two
particle graphs $\daleth$ (which indicates where the curvature of
spacetime is modified by the presence of matter), and $\daleth^*$,
(which describes the actual Feynman graph of the field and the
propagation of spin and angular momentum degrees of freedom), are
topologically equivalent and so necessitates a different
structure. It would be very interesting to know whether it is
possible to generalise the model of \cite{us} to spinning
particles, but this will only be the subject of future work.
%
%
%
%
%
%
%
%
%
%
\section{Conclusions and Outlook}
In this paper we have presented a group field theory formulation
of 3-dimensional Riemannian quantum gravity coupled to matter
fields of any mass and any spin, thus generalising the work of
\cite{us}; the model is a rather simple generalisation of the
Boulatov model for pure 3d gravity, and in particular simpler in
structure than the one presented in \cite{us}, despite the fact
that the configurations generated by the last arise as a particular
case of the new model. The new model reproduces exactly the spin
foam amplitudes for gravity coupled to particles constructed in
\cite{laurentPRI}, and can be seen as a simultaneous realisation
of a simplicial third quantization of gravity and a second
quantization of matter. In fact the perturbative expansion of the
partition function of the group field theory produces at once a
sum over 3d simplicial complexes of any topology, a sum over the
corresponding geometries, and a sum over Feynman graphs for matter
fields interactions.

Matter configurations arise as topological defects of gravity
configurations labeled by the Poincar\'e group, thus carrying mass
and spin, consistently with the known results for the quantization
of point particles in 3d \cite{DJT, matschullwelling}.

These results on the one hand confirm the flexibility and power of
the group field theory formalism, on the other hand give further
support to the view that it represents a fundamental definition of
quantum gravity in terms of spin foams and not merely an auxiliary
formalism.

\medskip
Most important, we believe that our results may be crucial for
further developments in this area. Let us then give a brief
outlook of possible future work, in relation to what we have
presented in this paper. An important achievement that was made
possible by the construction of a spin foam model for 3d quantum
gravity coupled to matter \cite{laurentPRI} was the identification
of an effective non-commutative field theory for (scalar) matter
that reproduces the Feynman amplitudes including the quantum
gravity corrections in the perturbative expansion
\cite{laurentPRIII}. The importance is also that this result on
the one hand clarifies the role of non-commutative geometry in
quantum gravity from the point of view of a fundamental
formulation of the theory, on the other hand it connects directly
and precisely spin foam models with effective models of quantum
gravity in flat spacetimes like Deformed (or Doubly) Special
Relativity \cite{jurek}, thus representing a good starting point
for tackling issues of quantum gravity phenomenology. Therefore,
having now obtained a group field theory that produces the Feynman
amplitudes of \cite{laurentPRI}, the first issue is to derive and
understand the non-commutative field theory of \cite{laurentPRIII}
and its extensions (e.g. to non-zero spin) from the group field
theory itself. It is natural to expect that it is the very action
of the new GFT we have constructed in this paper that, after
suitable integration over quantum gravity degrees of freedom, will
reduce to the effective non-commutative field theory for matter.
Work on this is indeed in progress \cite{usEffective}.

A second issue, to be tackled in the near future concerns gauge
fields. The model we presented can accommodate the description of
spin 1 fields with no difficulty but this is not enough to
interpret them as gauge bosons; first of all the case of zero mass
is not completely straightforward, and more work is needed to
understand it in full; second, and most important, the
interpretation of these interacting spin 1 particles as gauge
bosons for some (possibly non-abelian) gauge theory is not solid
at all. Work is in progress \cite{usGYM} on the construction of a
coupled and possibly unified model of quantum gravity and
Yang-Mills theory at the level of group field theory, inspired by
the results obtained in \cite{danhend} at the spin foam level, in
4 spacetime dimensions. However, the are two main obstacles in
constructing a complete group field theory in which all types of
matter fields, bosonic and fermionic, Yang-Mills fields and
quantum gravity are encoded in one action, combining the results
obtained in this paper, and those of \cite{usGYM}: the work of
\cite{danhend, usGYM} is in 4 dimensions and the corresponding
model in 3 dimensions is not easily constructed; most important,
in \cite{danhend,usGYM} Yang-Mills theory is obtained in a {\it
non-perturbative lattice formulation} (suitably generalised to
couple it with quantum gravity), while in the work we have
presented here we were able to reproduce the {\it perturbative}
interactions of bosonic particles in terms of Feynman diagrams; to
reconcile the two pictures is not straightforward, although it is
clearly possible.

Going back to issues related to group field theories in general,
of paramount importance is a complete understanding of gravity
symmetries, that can be nicely identified and taken care of at the
level of spin foam amplitudes \cite{laurentdiffeo}, at the level
of the group field theory itself, being it the classical action or
the partition function of the theory. In particular, the group
field theory origin and manifestation of the translation symmetry
characterizing $BF$ theories in any dimension and thus gravity in
3 dimensions is still unclear and must be studied as a matter of
priority. This is important because if spin foam symmetries are
not understood as symmetries of the corresponding group field
theories, it would be hard to maintain the latter as a fundamental
definition of the former; moreover, translation symmetries may be
the easiest context in which to develop techniques and ideas for
tackling the general issues of symmetries in group field theories
and for understanding the quantum origin of the classical
symmetries of gravity actions. Of course, diffeomorphism symmetry
is, in this respect, the ultimate target.

Needless to say, the ultimate goal of the work whose results we
presented in this paper is the issue of matter coupling to quantum
gravity in 4 spacetime dimensions, again for both a theoretical
interest and a move towards quantum gravity phenomenology. As
concerns this issue, the results obtained can turn out to be
useful in that they lend themselves to a straightforward
generalisation to higher dimensions, albeit a formal one; the
difficulty in fact is not so much the extension of techniques and
structures used here, to group field theory models of
4-dimensional quantum gravity, but the physical interpretation in
terms of matter fields of the resulting model. To understand
matter coupled to quantum gravity in 4 dimensions, one can start,
in a sense as it was done in 3 dimensions, from either classical
actions for gravity coupled to matter or from Feynman diagrams of
the matter quantum field theory \cite{laurentaristide}, then
construct the corresponding coupled spin foam models, and finally
obtain the group field theory formulation of them. Our results, if
suitably generalised to 4-dimensions, would allow to proceed the
other way around: start from a group field theory that gives 4d
quantum gravity as a spin foam model with extra structures that
can be hoped to represent matter, in the light of our results,
obtain the corresponding spin foam amplitudes, and either study
the no-gravity limit to understand the matter interpretation of
the resulting theory, or try to extract an effective
non-commutative field theory that admits such interpretation. This
is of course a longer term programme, that however is made a bit
easier by our results.

\section{Acknowledgements}
We would like to warmly thank J. W. Barrett, L. Freidel, K. Krasnov, E. Livine,
K. Noui and A. Perez for many helpful discussions.

\appendix
\section{Coupled Ponzano-Regge model}
\label{cpr} We outline briefly some characteristic properties of
the Coupled Ponzano-Regge model developed in \cite{laurentPRI}. To
begin, the model is a spin foam quantisation of first order 3d
Riemannian gravity coupled to spinning point particles.  One
discretises a 3-manifold $\mathcal{M}$ using a triangulation $\D$,
constituting of tetrahedra, triangles, edges ($e$), and vertices
($v$). The particle graph is discretised also in this process and
is replaced by a sequence of contiguous edges, $\daleth$,
contained in $\D$. No longer are the dynamical quantities
represented by entities which are continuous on the manifold, but
this information now resides in their discrete analogues.  We can
construct also the topological dual of $\D$, by placing a vertex
($v^*$) at the centre of every tetrahedron, and joining the
vertices of adjacent tetrahedra by edges ($e^*$) passing through
the centre of the triangles.  This is called the dual 1-skeleton
of the triangulation.  These dual edges form loops or faces
($f^*$) around the original edges of the triangulation.  These
dual faces together with the dual 1-skeleton form the dual
2-skeleton $\D^*$.  We label this structure with the discrete
dynamical variables and the quantum amplitude is a function of
these variables, it is a spin foam.  The action for our theory is
a BF action minimally coupled to the point particle action.  The
essential dynamical quantity for the gravity sector is the
holonomy, the parallel transport of the connection, along a dual
edge, $\a_{e^*}$. From this we form the discrete analogue of the
curvature.  As it is a 2-form, we discretise  it onto the dual
faces. Its discrete form is the holonomy around the dual edges
bounding $f^*$ denoted $G_e$. Remember that the edges $e$ of the
triangulation are in one-one correspondence with the faces $f^*$
of the dual so we use these notations interchangeably.  The point
particle in 3d is defined by its mass and spin. In the quantum
regime, these are encoded as the momentum, and a spin projector
both associated with $e$. The momentum is an $\SU(2)$ group
element $u_e^{-1}hu_e$ in the conjugacy class of a certain $\U(1)$
element $h$ encoding the mass of the particle.  To contain the
spin of the particle we associate to the edges of the particle
graph $\daleth$, two representations of $\SU(2)$, $I_e$ and
$I'_e$, one at each vertex.  to the edge itself we associate a
spin projector which we describe shortly.

The quantum amplitude for a manifold with particle graph is given
as follows.  For edges of the triangulation but not in the
particle graph $\daleth$, we assign to the edge a $\d$-function
over the holonomy of the associated dual face.  This is the usual
curvature flatness condition.  For edges of $\daleth$, the
discretised particle variables cause defects, and contribute to
the amplitude. The mass breaks the flatness condition and the
$\d$-function is now over the product of the holonomy and the
particle's momentum. Furthermore, the particle's spin contributes
a factor which may be visualised as a particle graph $\daleth^*$
in the dual.  The dual particle graph is a series of contiguous
edges in $\D^*$ which have the same topology as $\daleth$ and lie
\lq close' to $\daleth$.  We define the term \lq close' later.  We
associate to certain vertices of $\daleth^*$ total angular
momentum intertwiners.  We do this in the obvious way.  If a
trivalent interaction occurs in $\daleth$, then three total
angular momentum representations label that vertex.  By
topological equivalence there will be a trivalent vertex in
$\daleth^*$ and we label it with an intertwiner over the three
total angular momentum variables. Between the intertwiners we
place the matrix elements of the holonomy along the dual edges in
the total angular momentum representations $I_e$ and $I'_e$.  At
some point At the point where these holonomies meet we place the
spin projector
\begin{equation}
D^{I_e}_{.s}(u_e^{-1})D^{I'_e}_{s.}(u_e).
\end{equation}
For a more precise definition of where this happens exactly, see
\cite{laurentPRI}.  This completes the description of the
amplitude which we write down mathematically as
\begin{equation}
Z_{CPR}[\daleth]=\int
\prod_{e^*\in\D^*}d\a_{e^*}\prod_{e\in\D/\daleth}\d(G_e)\prod_{e\in\daleth}\int
du_e\; \d(G_eu_e^{-1}hu_e)
\sum_{I_e,I'_e}D^{I_e}_{.s}(x_e^{-1}u_e^{-1})
D^{I'_e}_{s.}(u_ex'_e) \times\left(\substack{\textrm{Total
angular}\\ \textrm{momemtum intertwiners}}\right),
\end{equation}
where $G_e$, $x_e$ and $x'_e$ are all products of the holonomies
$\a_{e^*}$.  We have only included a spin s particle but we could
include many more just by making the mass and spin dependent on
the edge, that is, $h\rar h_e$ and $s\rar s_e$.  This is a viable
proposition because these amplitudes have the properties of
implicit momentum conservation which we explain in detail in
Section \ref{ampdisc} and of implicit spin conservation, in that
we only intertwine the total angular momenta but the only non-zero
amplitudes are those for which the spin is conserved at the
interaction vertices.

An important property of the CPR spin foam model, is that the
particle graphs are topologically equivalent and also that they
are \lq close' together.  By close we mean that the dual particle
graph $\daleth^*$ lies only on those edges $e^*$ of the spin foam
$\D^*$ lying in the dual tube $T_{\daleth}^*$, where
\begin{equation}\begin{split}T_{\daleth}^*&=\left\{f^*\in\D^*:\quad f^* \;\; \textrm{dual to}
\;\; e\in T_{\daleth} \subset \D  \right\},\\
T_{\daleth}&=\left\{e\in\D:\quad
e\cap\daleth\neq0,e\notin\daleth\right\}.
\end{split}\end{equation}
In words, the dual tube is the set of faces $f^*$ which are dual
to edges $e$ of the triangulation that share a vertex with the
particle graph $\daleth$ but are not in $\daleth$.  These required
properties of the two graphs are satisfied by our gft.

\section{Mode expansion of the fields}
\label{exp} We do some calculations relating to the kinematic
regime of our model.  We perform a Peter-Weyl decomposition of the
Boulatov field into its constituent representations.
\begin{equation}P_{\a}\p(g_1,g_2,g_3)= \sum_{\substack{j_i,m_i,n_i,k_i\\1\leq i\leq
3}}\p^{j_1j_2j_3}_{m_1k_1m_2k_2m_3k_3} D^{j_1}_{m_1n_1}(g_1)
D^{j_2}_{m_2n_2}(g_2) D^{j_3}_{m_3n_3}(g_3)\int d\a\;
D^{j_1}_{n_1k_1}(\a) D^{j_2}_{n_2k_2}(\a) D^{j_3}_{n_3k_3}(\a)
\end{equation}
But the following equality holds
\begin{equation}\int d\a\;
D^{j_1}_{k_1n_1}(\a) D^{j_2}_{k_2n_2}(\a) D^{j_3}_{k_3n_3}(\a) =
C^{j_1j_2\,j_3}_{k_1k_2k_3}C^{j_1\;j_2\,j_3}_{n_1n_2n_3},\end{equation}
where $C$ is an $\SU(2)$ trivalent intertwiner.  So we define
\begin{equation}\P^{j_1\;\;j_2\;\;j_3}_{m_1m_2m_3}=\frac{1}{\sqrt{d_{j_1}d_{j_2}d_{j_3}}}
\p^{j_1j_2j_3}_{m_1k_1m_2k_2m_3k_3},
\end{equation}
and this means that we can write the above projected field as in
equation (\ref{pexp}):
\begin{equation}P_{\a}\p(g_1,g_2,g_3)=\sum_{\substack{j_i,m_i,n_i \\ 1\leq i\leq
3}}\sqrt{d_{j_1}d_{j_2}d_{j_3}}\P^{j_1\;\;j_2\;\;j_3}_{m_1m_2m_3}\rep{1}\rep{2}\rep{3}C^{j_1\;j_2\,j_3}_{n_1n_2n_3}\end{equation}

We follow a the same procedure for the coupled field.  We expand
into representations
\begin{equation}\begin{split}P_{\a}\psi(g_1,g_2,g_3;u)=&\sum_{I,n,k}\quad\sum_{\substack{ j_i,m_i,n_i,k_i\\1\leq i\leq
3}}\psi^{j_1j_2j_3I}_{m_1k_1m_2k_2m_3k_3sk} D^{j_1}_{m_1n_1}(g_1)
D^{j_2}_{m_2n_2}(g_2) D^{j_3}_{m_3n_3}(g_3) D^{I}_{sn}(u)\\
&\hphantom{xxxxxxxxxxxxxxx}\times\int d\a\; D^{j_1}_{n_1k_1}(\a)
D^{j_2}_{n_2k_2}(\a) D^{j_3}_{n_3k_3}(\a) D^{I}_{nk}(\a).
\end{split}\end{equation} There is a similar equality that holds for a
product of four $\SU(2)$ representations: \begin{equation}\int
d\a\; D^{j_1}_{n_1k_1}(\a) D^{j_2}_{n_2k_2}(\a)
D^{j_3}_{n_3k_3}(\a)
D^{I}_{nk}(\a)=\sum_{\L}\tilde{C}^{j_1\;j_2\,j_3\,I\,\L}_{n_1n_2n_3n}\tilde{C}^{j_1\;j_2\,j_3\,I\,\L}_{k_1k_2k_3k},\end{equation}
where $\tilde{C}$ is a 4-valent $\SU(2)$ intertwiner and $\L$
labels a basis in the vector space of intertwiners. Once again we
define
\begin{equation}\Psi^{j_1\;\;j_2\;\;j_3\;\,I\;\L}_{m_1m_2m_3s}=\frac{1}{\sqrt{d_{j_1}d_{j_2}d_{j_3}d_{I}}}\psi^{j_1j_2j_3I}_{m_1k_1m_2k_2m_3k_3sk},\end{equation}
and so we can write our field as in (\ref{cexp}):
\begin{equation}P_{\a}\psi_s(g_1,g_2,g_3;u)=\sum_{\substack{I,n,j_i,m_i,n_i\\
1\leq i\leq
3}}\sum_{\L}\sqrt{d_{j_1}d_{j_2}d_{j_3}d_I}\Psi^{j_1\;\;j_2\;\;j_3\;\,I\;\L}_{m_1m_2m_3s}
\rep{1}\rep{2}\rep{3}D^{I}_{sn}(u)\tilde{C}^{j_1\;j_2\,j_3\,I\,\L}_{n_1n_2n_3n}
\end{equation}

\section{1-4 Equality}
\label{onefour} The 1-4 equality was given pictorially as FIG.
\ref{2teteq} in Section \ref{cdyn}.  We prove this statement here
for the vertex term with a bivalent particle interaction but it
hold for the trivalent term also. We couch the proof of the
equality in terms of the action term as this has a self-contained
integration over all variables and allows us to maintain control
and knowledge of all redefinitions of the variables. The lhs of
FIG. \ref{2teteq} has the action term as stated in
(\ref{actionexp}) with the operator (\ref{vc2})
\begin{equation}\begin{split}&\int\prod^{6}_{i=1}dg_i\,d\bar{g}_i\, du_a\, du_b\;
\psi_{s}(g_1,g_2,g_3;u_a)\psi_{s}(g_4,\bar{g}_3,g_5;u_b)\p(\bar{g}_4,\bar{g}_2,g_6)\p(\bar{g}_6,\bar{g}_5,\bar{g}_1)\\
&\hphantom{xxxx}\times\mu_2\int\prod_{i=1}^{4}d\a_i\;
\dvp{1}{4}{1}\dvp{2}{3}{1}\dvc{3}{2}{1}{a}\\
&\hphantom{xxxxxxxxxxxxxxxxxxxx}\times\dvc{4}{3}{2}{b}\dvp{5}{4}{2}\dvp{6}{4}{3}\\
&\hphantom{xxxxxxxxxxxxxxxxxxxxxxxxxxxxx}\times\d(\a_1u_a^{-1}hu_a\a_1^{-1}\a_2u_b^{-1}h^{-1}u_b\a_2^{-1})
\sum_{n_a,n_b}D^{I}_{sn_a}(u_a\a_1^{-1})D^{I}_{sn_b}(u_b\a_2^{-1})\d_{n_an_b}.
\label{biv}\end{split}\end{equation} Now we start from the other
end. The vertex term for the rhs of FIG. \ref{2teteq} is
\begin{equation}\begin{split}&\int\prod^{6}_{i=1}dg_i\,d\bar{g}_i\, du_a\,
du_b\int\prod_{i=1}^{4}d\a_i\prod_{j,k =1\; :\; j<k}^{4}
d\g_{jk}\;
\psi_{s}(g_1,g_2,g_3;u_a\g_{14}\a_1)\psi_{s}(g_4,\bar{g}_3,g_5;u_b\g_{24}\a_2)\p(\bar{g}_4,\bar{g}_2,g_6)\p(\bar{g}_6,\bar{g}_5,\bar{g}_1)\\
&\hphantom{xxxx}\times\mu_2
\dvg{1}{4}{1}\dvg{2}{3}{1}\dvg{3}{2}{1}\dvg{4}{3}{2}\dvg{5}{4}{2}\dvg{6}{4}{3}\\
&\hphantom{xxxxxxxxxxxxxx}\times\d(\g_{12}^{-1}\g_{24}^{-1}u_a^{-1}h^{-1}u_a\g_{14})\d(\g_{13}^{-1}\g_{23}\g_{12})\d(\g_{23}^{-1}\g_{34}^{-1}u_b^{-1}hu_b\g_{24})\d(\g_{13}\g_{14}^{-1}\g_{34})\\
&\hphantom{xxxxxxxxxxxxxxxxxxxxx}\sum_{n_a,n_b}D^{I}_{sn_a}(u_a)D^{I}_{sn_b}(u_b)\d_{n_an_b}.
\end{split}\end{equation}
Upon integrating with respect to $\g_{12}$, $\g_{13}$ and
$\g_{23}$ our action reduces to
\begin{equation}\begin{split}&\int\prod^{6}_{i=1}dg_i\,d\bar{g}_i\, du_a\,
du_b\int\prod_{i=1}^{4}d\a_i\prod_{j=1}^{3} d\g_{j4}\;
\psi_{s}(g_1,g_2,g_3;u_a\g_{14}\a_1)\psi_{s}(g_4,\bar{g}_3,g_5;u_b\g_{24}\a_2)\p(\bar{g}_4,\bar{g}_2,g_6)\p(\bar{g}_6,\bar{g}_5,\bar{g}_1)\\
&\hphantom{xxxx}\times\mu_2\dvg{1}{4}{1}\d(\bar{g}_2\a_3^{-1}\g_{34}^{-1}\g_{14}\a_1g_2^{-1})\d(\bar{g}_3\a_2^{-1}\g_{24}^{-1}u_a^{-1}h^{-1}u_a\g_{14}\a_1g_3^{-1})\\
&\hphantom{xxxxxxxxxxxx}\dvg{4}{3}{2}\d(\bar{g}_5\a_3^{-1}\g_{34}^{-1}u_b^{-1}hu_b\a_{2}g_5^{-1})\dvg{6}{4}{3}\\
&\hphantom{xxxxxxxxxxxxxxxxx}\times\d(u_a^{-1}hu_au_b^{-1}h^{-1}u_b)\sum_{n_a,n_b}D^{I}_{sn_a}(u_a\a_1^{-1})D^{I}_{sn_b}(u_b\a_2^{-1})\d_{n_an_b}.
\end{split}\end{equation}
and redefining $\g_{14}\a_{1}\rar\a_{1}$,
$\g_{24}\a_{2}\rar\a_{2}$ and $\g_{34}\a_{3}\rar\a_{3}$ followed
by $u_a\a_1\rar u_a$ and $u_b\a_2\rar u_b$ we end up with
(\ref{biv}).

\section{2-3 Equality}
\label{twothreesect} We drew the 2-3 move in FIG. \ref{twothree}
of Section \ref{qdyn}.  Here, we will prove this relation
explicitly. We place the amplitudes for the lhs and rhs figures in
two columns:
\begin{displaymath}
\begin{array}{c|c}
\includegraphics[width=4cm]{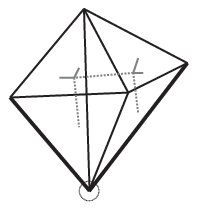} &
\includegraphics[width=4cm]{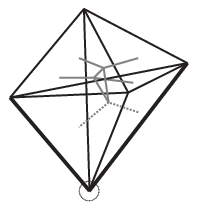}\\ & \\ & \\ & \\
\d(g'_1\ba_1^{-1}\ba_4\a_4^{-1}\a_1g_1^{-1}) &
\d(g'_1\g_1^{-1}\g_2g_1^{-1})\\
\d(g'_5\ba_2^{-1}\ba_4\a_4^{-1}\a_2g_5^{-1}) &
\d(g'_5\tg_1^{-1}\tg_2g_5^{-1})\\
\d(g'_9\ba_3^{-1}\ba_4\a_4^{-1}\a_3g_9^{-1}) &
\d(g'_9\bg_1^{-1}\bg_2g_9^{-1})\\ & \\
\d(g'_2\ba_1^{-1}u_a^{-1}hu_a\ba_3g{'}_8^{-1}) &
\d(g'_2\g_1^{-1}\g_4u_a^{-1}hu_a\bg_3^{-1}\bg_1g{'}_8^{-1})\\
\d(g_2\a_1^{-1}u_b^{-1}h^{-1}u_b\a_3g_8^{-1}) &
\d(g_2\g_2^{-1}\g_4u_b^{-1}h^{-1}u_b\bg_3^{-1}\bg_2g_8^{-1})\\ &
\\ \d(g'_3\ba_1^{-1}\ba_2g{'}_4^{-1}) &
\d(g'_3\g_1^{-1}\g_3\tg_4^{-1}\tg_1g{'}_4^{-1})\\
\d(g_3\a_1^{-1}\a_2g_4^{-1}) &
\d(g_3\g_2^{-1}\g_3\tg_4^{-1}\tg_2g_4^{-1})\\ & \\
\d(g'_6\ba_2^{-1}\ba_3g{'}_7^{-1}) &
\d(g'_6\tg_1^{-1}\tg_3\bg_4^{-1}\bg_1g{'}_7^{-1})\\
\d(g_6\a_2^{-1}\a_3g_8^{-1}) &
\d(g_6\tg_2^{-1}\tg_3\bg_4^{-1}\bg_2g_7^{-1})\\ & \\
 & \d(\g_3^{-1}\g_4\bg_3^{-1}\bg_4\tg_3^{-1}\tg_4)\\ & \\
 \d(u_a\a_4\ba_4^{-1}u_b^{-1}) & \d(u_au_b^{-1})\\
 & \\
 D^{I_a}_{s.}(u_a\a_3)D^{s}_{sm}(u_a\a_4)\;\; D^{s}_{sm}(u_b\ba_4)D^{I_a}_{s.}(u_b\ba_3)\quad
 &\quad
 D^{I_a}_{s.}(u_a\bg_2)D^{s}_{sm}(u_a)\;\; D^{s}_{sm}(u_b) D^{I_a}_{s.}(u_b\bg_1)
\end{array}
\end{displaymath}
From here we are going to manipulate the rhs to have the same form
as the left by redefining variables and using one integration. We
have neglected to insert the explicit integration of the
variables.
\begin{itemize}
\item Step 1. Redefine the following variables for the rhs:
\begin{equation}\bar{\g}_3^{-1}\bar{\g}_1\rar\bar{\g}_1,\quad
\bar{\g}_3^{-1}\bar{\g}_2\rar\bar{\g}_2,\quad
{\g}_4^{-1}{\g}_1\rar{\g}_1, {\g}_4^{-1}{\g}_2\rar{\g}_2,\quad
\tilde{\g}_4^{-1}\tilde{\g}_1\rar\tilde{\g}_1,\quad
\tilde{\g}_4^{-1}\tilde{\g}_2\rar\tilde{\g}_2.\end{equation}
\item Step 2. Integrate w.r.t. $\tilde{\g}_4$.\\
\item Step 3. Redefine:
$\quad\g_4^{-1}\g_3\tilde{\g}_1\rar\tilde{\g}_1,\quad
\g_4^{-1}\g_3\tilde{\g}_2\rar\tilde{\g}_2$. At this point the rhs
looks like:
\begin{equation}\begin{split}&\d(g'_1\g_1^{-1}\g_2g_1^{-1})\d(g'_5\tg_1^{-1}\tg_2g_5^{-1})\d(g'_9\bg_1^{-1}\bg_2g_9^{-1})\\
&\d(g'_2\g_1^{-1}u_a^{-1}hu_a\bg_1g{'}_8^{-1})\d(g_2\g_2^{-1}u_b^{-1}h^{-1}u_b\bg_2g_8^{-1})\\
&\d(g'_3\g_1^{-1}\tg_1g{'}_4^{-1})\d(g_3\g_2^{-1}\tg_2g_4^{-1})\\
&\d(g'_6\tg_1^{-1}\bg_1g{'}_7^{-1})\d(g_6\tg_2^{-1}\bg_2g_7^{-1})\\
&\d(u_au_b^{-1}) D^{I_a}_{s.}(u_a\bg_2)D^{s}_{sm}(u_a)\;\;
D^{s}_{sm}(u_b) D^{I_a}_{s.}(u_b\bg_1)\end{split}
\end{equation}

\item Step 4. Relabel:  $\quad\g_1\rar\ba_4^{-1}\ba_1,\quad\tg_1\rar\ba_4^{-1}\ba_2,\quad\bg_1\rar\ba_4^{-1}\ba_3,
\quad\g_2\rar\a_4^{-1}\a_1,\quad\tg_2\rar\a_4^{-1}\a_2,\quad\bg_2\rar\a_4^{-1}\a_3.$\\
\item Step 5. Redefine: $\quad u_a\rar u_a\a_4,\quad u_b\rar
u_b\ba_4.$
\end{itemize}
That finishes the proof of the equality.

\section{Equality of the two bivalent vertex terms}
\label{simp} In Section \ref{cdyn}, equation (\ref{biveq}), we
stated a result concerning the equality of two vertex terms with a
bivalent particle interaction. We prove this here. The new term as
given in (\ref{biveq}) is
\begin{equation}\begin{split}\mu_2\int \prod^{6}_{i=1}dg_i\, du_a\, du_b\;
&P_{\a_1}\psi_{s}(g_1,g_2,g_3u_a^{-1}hu_a;u_a)P_{\a_2}\psi_{s}(g_4u_b^{-1}h^{-1}u_b,g_3,g_5;u_b)
P_{\a_3}\p(g_4,g_2,g_6)P_{\a_4}\p(g_6,g_5,g_1)\\
&\phantom{xxxxxxxxxx}\times
\d(u_a^{-1}hu_au_b^{-1}h^{-1}u_b)\sum_{n_a,n_b}D^{I}_{sn_a}(u_a)D^{I}_{sn_b}(u_b)\d_{n_an_b}\end{split}\end{equation}
Upon integrating with respect to $u_b$ we find that the
$\d$-function is satisfied if $u_b=ku_a$ for all $k\in U(1)$ the
same $U(1)$ subgroup that contains $h$.  Thus our vertex becomes
\begin{equation}\begin{split}\mu_2\int \prod^{6}_{i=1}dg_i\, du_a\,
du_b \int_{U(1)} dk\;
&P_{\a_1}\psi_{s}(g_1,g_2,g_3u_a^{-1}hu_a;u_a)P_{\a_2}\psi_{s}(g_4u_a^{-1}h^{-1}u_a,g_3,g_5;ku_a)
P_{\a_3}\p(g_4,g_2,g_6)\\ &\phantom{xxxxxxxxxx}\times
P_{\a_4}\p(g_6,g_5,g_1)D^{I}_{ss}(k)\end{split}\end{equation} But
we note that $D^{J}_{ss}=e^{-is\o_k}$ for all $J$ and a given
$\o_k$ such that $0\leq \o_k\leq 2\pi$. Thus,
$\psi_{s}(g_4u_a^{-1}h^{-1}u_a,g_3,g_5;ku_a)=e^{is\o_k}\psi_{s}(g_4u_a^{-1}h^{-1}u_a,g_3,g_5;ku_a)$
and the $k$ dependent part of this amplitude is
\begin{equation}\int^{2\pi}_{0} \frac{d\o_k}{2\pi} e^{-is\o_k}e^{is\o_k}=1.\end{equation}
We are finally left with \begin{equation}\mu_2\int
\prod^{6}_{i=1}dg_i\, du_a\;
P_{\a_1}\psi_{s}(g_1,g_2,g_3u_a^{-1}hu_a;u_a)P_{\a_2}\psi_{s}(g_4u_a^{-1}h^{-1}u_a,g_3,g_5;u_a)
P_{\a_3}\p(g_4,g_2,g_6)P_{\a_4}\p(g_6,g_5,g_1),\end{equation}
which is independent of $k$ and $I$ as promised.

\end{document}